\documentclass[twocolumn]{svjour3}          
\smartqed  
\usepackage{graphicx}
\usepackage{amsmath}
\usepackage{nicefrac}
\usepackage{amssymb}
\usepackage{epsfig}
\usepackage{url}
\usepackage[at]{easylist}

    \newtheorem{prop}{Proposition}
    \newtheorem{lem}[prop]{Lemma}

    \newtheorem{thm}{Theorem}

\newcommand{\Painleve}{\mbox{Painlev\'{e} $\;$}}
\newcommand{\Painleves}{\mbox{Painlev\'{e}'s }}

\newtheorem{defn}{Definition}
\newcommand{\seq}{\mbox{$\! \, = \, \!$}}

\newcommand{\Ag}{\mbox{$\alpha$}}

\newcommand{\kg}{\mbox{$\kappa$}}

\newcommand{\tg}{\mbox{$\theta$}}
\newcommand{\td}{\mbox{$\dot \theta$}}
\newcommand{\tdd}{\mbox{$\ddot \theta$}}

\newcommand{\vq}{\mbox{${\mathbf{q}}$}}

\newcommand{\vfh}{\mbox{${\mathbf{\hat f}}$}}

\newcommand{\vf}{\mbox{${\mathbf{f}}$}}
\newcommand{\vfi}{\mbox{$\varphi$}}
\newcommand{\vn}{\mbox{${\mathbf{n}}$}}
\newcommand{\vp}{\mbox{${\mathbf{p}}$}}
\newcommand{\vt}{\mbox{${\mathbf{t}}$}}

\newcommand{\vR}{\mbox{${\mathbf{R}}$}}

\newcommand{\vJ}{\mbox{${\mathbf{J}}$}}
\newcommand{\vP}{\mbox{${\mathbf{P}}$}}

\newcommand{\vr}{\mbox{${\mathbf{r}}$}}

\newcommand{\vy}{\mbox{${\mathbf{y}}$}}

\newcommand{\vA}{\mbox{${\mathbf{A}}$}}

\newcommand{\Ss}{\mbox{${\mathcal{S}}$}}

\newcommand{\vqd}{\mbox{${\mathbf{ \dot q}}$}}

\newcommand{\vrd}{\mbox{${\mathbf{ \dot r}}$}}
\newcommand{\vrdd}{\mbox{${\mathbf{ \ddot r}}$}}

\newcommand{\vRdd}{\mbox{${\mathbf{ \ddot R}}$}}

\newcommand{\xd}{\mbox{$\dot{x}$}}
\newcommand{\xdd}{\mbox{$\ddot{x}$}}

\newcommand{\zd}{\mbox{$\dot{z}$}}
\newcommand{\zdd}{\mbox{$\ddot{z}$}}

\newcommand{\real}{\mbox{$\mathbb{R}$}}
\newcommand{\Fs}{\mbox{$\cal F$}}
\newcommand{\As}{\mbox{$\cal A$}}

\newcommand{\Poincare}{Poincar\'{e} }

\newcommand{ \matwo }[4] {
 \left( \begin{array}{c c }
   #1  &  #2     \\
   #3  &  #4
\end{array} \right)  }
\newcommand{\beq}[1]{ \begin{equation} \label{eq.#1} }
\newcommand{\eeq}{ \end{equation} }
\newcommand{\eq}[1]{(\ref{eq.#1})}

\DeclareMathOperator{\arccot}{arccot}
\begin{document}

\title{Lyapunov stability of a rigid body with two frictional contacts \thanks{PLV acknowledges support from
the National Research, Innovation and Development Office of Hungary under grant K104501.}
}


\author{P\'eter L. V\'arkonyi \and Yizhar Or.
}


\institute{P. L. V\'arkonyi \at
              Dept. of Mechanics Materials and Structures, Budapest University of Technology and Economics, H-1111 Budapest, Hungary\\
              Tel.: +36 1 463 1317\\
              Fax: +36 1 463 1773\\
              \email{vpeter@mit.bme.hu}           
           \and
           Y. Or \at
              Faculty of  Mechanical Engineering,
Technion - Israel Institute of  Technology, Haifa 3200003, Israel \\
              \email{izi@technion.ac.il}
}

\date{Received: date / Accepted: date}

\maketitle

\begin{abstract}
Lyapunov stability of a mechanical system means that the dynamic response stays bounded in an arbitrarily small neighborhood of a static equilibrium configuration under small perturbations in positions and velocities. This type of stability is highly desired in robotic applications that involve multiple unilateral contacts. Nevertheless, Lyapunov stability analysis of such systems is extremely difficult, because even small perturbations may result in hybrid dynamics where the solution involves many nonsmooth transitions between different contact states.  This paper concerns with Lyapunov stability analysis of a planar rigid body with two frictional unilateral contacts under inelastic impacts, for a general class of equilibrium configurations under a constant external load. The hybrid dynamics of the system under contact transitions and impacts is formulated, and a \Poincare map at two-contact states is introduced. Using invariance relations, this \Poincare map is reduced into two semi-analytic scalar functions that entirely encode the dynamic behavior of solutions under any small initial perturbation. These two functions enable determination of Lyapunov stability or instability for almost any equilibrium state. The results are demonstrated via simulation examples and by plotting stability and instability regions in two-dimensional parameter spaces that describe the contact geometry and external load.
\end{abstract}

\section{Introduction}

Lyapunov stability of an equilibrium state is a fundamental concept in dynamical systems theory \cite{leine.stability,lyapunov_book}. For mechanical systems, it means that the dynamic response stays bounded in a small neighborhood of a static equilibrium configuration under small perturbations in the system's state, i.e. positions and velocities. This type of stability is highly desired in robotic applications such as grasping, quasistatic manipulation and legged locomotion, which commonly involve intermittent contacts.

Dynamic stability of multi-contact equilibrium postures has been analyzed mainly under the assumption of compliant contacts, where the source of compliance is either small elastic deformations of contacting material surfaces \cite{howard&kumar,shapiro_passive_FC_ijrr2010,szalai2014prsa}, or force feedback of robotic fingers in force-closure grasps \cite{fen96,nguyen89,varkonyi2015stability}. These works, which typically analyze potential elastic energy of contact forces, assume that all contacts are maintained without separation or slippage, which is not always the case in practical scenarios. When one considers unilateral contact constraints of rigid bodies under friction bounds, the stability problem becomes much more involved, and most of the works resort to weaker definitions of `static stability', such as force closure grasps in robotic manipulation \cite{ponce&merlet_ijrr} which require existence of equilibrating reaction forces under frictional constraints, the ZMP criterion in legged locomotion \cite{ponce&merlet_ijrr} which checks that the action line of the resultant contact force does not exceed beyond the contact surface, or stability margins for rough-terrain vehicles such as the amount of initial energy required for tipover beyond basin of attraction of a local minimum of potential energy \cite{papadopoulos_icra96}. Another alternative approach is modelling the contact forces and stick-slip transition using the notion of Filippov systems \cite{champneys.book,filippov}, but this does not account for nonsmooth transitions due to contact separation and impacts.

In (multi-) rigid body systems under unilateral frictional contacts,
analysis of the dynamic response in the vicinity of an equilibrium state is a challenging task since it requires consideration of  various transitions between different contact states, including separation, impacts and stick-slip transitions \cite{stewart_siam00}. Such systems can be formulated as hybrid dynamical systems \cite{back1993hybrid,teel_tutorial09} or alternatively, as complementarity systems \cite{anitescu97,trinkle.comp}. The solution undergoes discrete transitions between contact modes, and its analysis may suffer from difficulties such as solution indeterminacy or inconsistency due to \Painleve paradoxes \cite{champneys&varkonyi_survey2016,genot99,leine2002,mason&wang,varkonyi.painleve2c.2016}, as well as dynamic jamming phenomenon of finite-time divergence \cite{or.RCD2014,or&rimon.NODY2012}.
Several works utilize complementarity formulation in order to study solution well-posedness of frictional multi-contact systems \cite{ballard2005,stewart_siam00}. The recent work \cite{blumentals2016} provides upper bounds on friction coefficients that guarantee avoidance of such Painlev\'{e}-type paradoxes. Another related criterion is
{\em strong stability} proposed in \cite{trinkle_zamm00}, which eliminates solution indeterminacy by requiring that static equilibrium is not ambiguous with other non-static contact solutions. 
Despite its name, this criterion is not directly related to Lyapunov stability since it does not consider solutions of the hybrid dynamics in response to local state perturbations.

Under small perturbations that involve contact separation, the dynamic response often undergoes {\em  impact} events due to collisions, which induce discontinuous velocity jumps. The simplest model of impact uses a kinematic coefficient of restitution and implicitly assumes frictionless contact impulse, which results in the famous class of bouncing-ball problems \cite{luck1993bouncing}. An important scenario in such cases is {\em Zeno behavior} \cite{or&ames_tac2011,zhang.zeno.hs}, where the solution undergoes an infinite sequence of exponentially decaying impacts that converge back to sustained contact in a finite bounded time. Several works analyzed the stability of so-called Zeno equilibrium states of single-contact systems by using approaches of Lyapunov functions \cite{goebel.cdc08.lyap.zeno,lamperski2013lyapunov,leine2012global} as well as \Poincare map analysis in which the discrete-time dynamics of impact states is investigated \cite{or&ames_tac2011,or&teel_TAC2011}.

Another related type of stability is {\em orbital stability} of hybrid periodic solutions that involve impacts. This issue has been studied extensively in the robotics literature on the control of dynamic legged locomotion \cite{hurmuzulu.biped04,biped_book_grizzle}. These works typically make the simplifying assumption of {\em perfectly inelastic impacts}, for which the  contact sticks immediately after collisions. This implies significant simplification of the stability analysis via linearization of the \Poincare map \cite{morris.tac2009}, since local perturbations always result in a solution with the same sequence of contact transitions. Almost all these works do not consider the possibility of slippage under friction limitations (except for \cite{gamus&or.siads2014,tavakoli2013}). Moreover, they do not consider the case of multiple contacts and transitions between multiple contact modes, which are a necessary component in stability analysis of multi-contact equilibrium point, even under arbitrarily small perturbations.


As for more realistic impact modelling under frictional contacts, several works on impact mechanics \cite{chatterjee&ruina_1998,stronge_book,mason&wang_impact} propose different laws of single-contact impact, which are primarily based on various definitions of restitution coefficients. The problem becomes even more complicated when considering impacts under multiple contacts \cite{brogliato,glocker1995multiple,ivanov1997impact}, as in the classical rocking block problem \cite{brogliato2012rocking,dimitrakopoulos2012revisiting}. A major simplification can be achieved if one assumes perfectly inelastic impacts, though friction constraints, slippage and multi-contact collisions may still pose difficulties. The work \cite{leine.NODY2008} studied dynamic stability of continuum equilibrium sets of planar mechanical systems under multiple frictional contacts, assuming impact laws under kinematic restitution coefficients. Choosing total mechanical energy as a Lyapunov function and using techniques of convex analysis and measure differential inclusions, explicit stability conditions were derived in \cite{leine.NODY2008}. These condition guarantee that the solution converges to an equilibrium set while staying close to a local minimum of the potential energy within the equilibrium set.
This is fundamentally different from our present work studying the Lyapunov stability of a specific equilibrium point embedded within a continuous set of equilibria, where the solution is required to stay within a bounded neighborhood of that particular point, while a minimum of potential energy may not exist at all. The work \cite{Basseville2003}  classified all possible equilibrium states of a point mass connected to two linear springs under unilateral frictional contact, and conducted Lyapunov stability analysis based on numerical simulations. Unlike in the present work, the definition of Lyapunov stability in \cite{Basseville2003} did not include finite-time convergence, since that model also considered elastic-induced equilibrium points without contacts which locally behave as a spring-mass oscillator, and neither display asymptotic nor finite-time convergence.

The recent work \cite{posa.stability.tac2016} has introduced an efficient computational algorithm that utilizes convex optimization for constructing sum-of-squares Lyapunov functions for planar mechanical systems with unilateral frictional contacts under inelastic impacts. This algorithm enables determination of Lyapunov stability for equilibrium \linebreak states, and even computation of conservative bounds on regions of attraction. While this efficient computational scheme is especially useful for feedback control design, it does not provide intuition on the physical mechanisms that determine stability or instability of uncontrolled contact systems. Moreover, this method is based on Lyapunov functions and does not provide conditions for {\em instability}, which are also important.

Two different mechanisms of instability have been identified in previous works that studied Lyapunov stability of equilibrium postures for a minimal model of a planar rigid body with two frictional contacts, which is similar to the classical problem of a rocking block \cite{dimitrakopoulos2012revisiting}. In the work \cite{or&rimon.icra08a}, it has been shown that one possible instability mechanism of frictional equilibrium postures can arise from solution indeterminacy. That is, \cite{or&rimon.icra08a} established that the strong stability criterion of \cite{trinkle_zamm00} is indeed a necessary condition for Lyapunov stability. The work \cite{Varkonyi.icra2012} identified a different destabilizing mechanism called {\em reverse chatter} \cite{nordmark.ima2010}, where the contact undergoes an exponentially diverging sequence of cyclic contact transitions and impacts under arbitrarily small initial perturbations. The work \cite{Varkonyi.icra2012} also provided conservative stability conditions by using an energy-based Lyapunov functions while assuming frictional impacts with tangential and normal coefficients of restitution, for a planar rigid body with two contacts on a slope. The work \cite{or&rimon.icra08b} considered arbitrary two-contact geometries while assuming frictionless impacts with normal restitution, and derived conservative conditions for Lyapunov stability of equilibrium postures by analyzing the \Poincare map of post-impact states. The two works \cite{Varkonyi.icra2012} and \cite{or&rimon.icra08b}  substantially differ in the choice of impact laws and in the techniques of stability proof. Nevertheless, both works chose to approximate the dynamics in a small neighborhood of equilibrium by assuming constant accelerations and contact forces for each possible contact mode. This concept, called {\em zero order dynamics}, has also been used in previous works on stability of single-contact Zeno equilibria, where value of the vector field of the dynamical system is held constant as evaluated at the nominal equilibrium point. 
Finally, the major shortcoming of \cite{Varkonyi.icra2012} and \cite{or&rimon.icra08b} is that they did not provide a sharp condition that determines whether a given frictional equilibrium configuration is Lyapunov stable or unstable, and leave a huge range of configurations with undecided stability classification.

\begin{figure*}[t]
\centering{\includegraphics[width=0.7\textwidth]{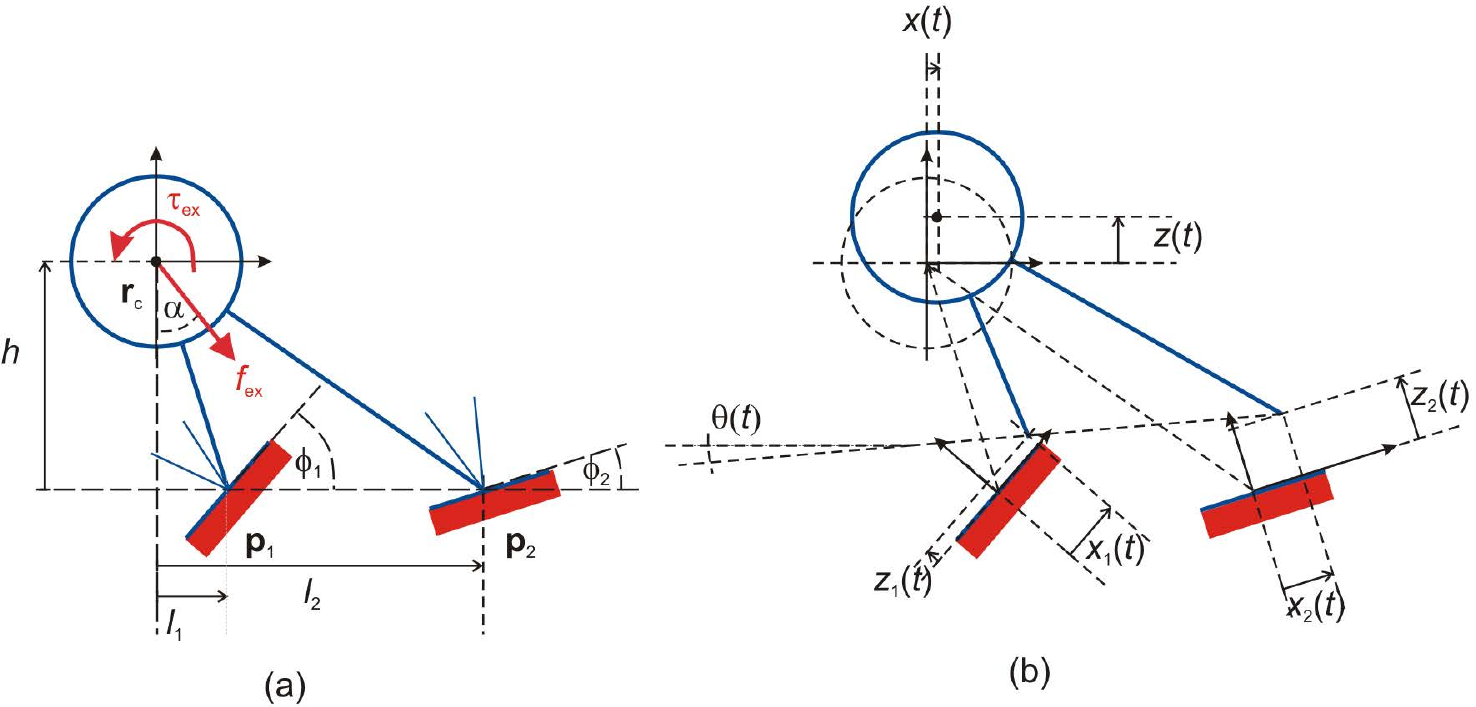}
\caption{The body on two frictional contacts - notation.}} \label{fig1}
\end{figure*}

The goal of this paper is to complement the two previous works \cite{or&rimon.icra08b,Varkonyi.icra2012} as well as \cite{posa.stability.tac2016} by presenting a semi-analytic method for determination of Lyapunov stability and instability of equilibrium configurations for a planar rigid body on two frictional contacts under a constant external load and inelastic impacts. We analyze all possible contact mode transitions under the zero-order dynamics approximation of the system, and define a generic class of equilibrium configurations called {\em persistent}, for which transitions from slipping contacts back to contact separation cannot occur. Then we introduce a \Poincare section that involves states of impact at one contact point and sustained contact at the other, along with its associated three-dimensional \Poincare map. Exploiting invariance relations, this \Poincare map is then reduced into two scalar maps which are both semi-analytic, and  determine the entire dynamic behavior of the system under any small initial perturbation. Using these two reduction maps, we identify simple sufficient conditions for stability or instability of an equilibrium configuration due to decaying or diverging Zeno sequences of contact transitions and impacts. Next, we present more general conditions that enable determination of stability for almost any persistent equilibrium configuration, by analyzing the interval graph structure of the reduced \Poincare map. The results are demonstrated by simulation examples and by plotting regions of stability and instability in  two-dimensional parameter spaces that describe the contact geometry and external load.

The organization of the paper is as follows. The next section introduces the problem statement, establishes notation and terminology, and defines the notion of finite-time Lyapunov stability. Section 3 considers all possible contact modes of the continuous-time dynamics, introduces the approximation of zero-order dynamics and the notion of solution ambiguity, and establishes that an ambiguous equilibrium is necessarily unstable. Section 4 introduces impact laws for single-contact and double-contact frictional inelastic collisions, and Section 5 presents the graph of all possible hybrid transition of solutions under the zero-order dynamics. Section 6 introduces the \Poincare map and its reduction map, and demonstrates how these maps relate to all possible behaviors of the dynamics via a series of examples. Section 7 presents the main contribution of the paper - conditions for Lyapunov stability and instability of two-contact frictional equilibrium configurations. The closing section briefly summarizes and discusses the results and their limitations, and the Appendix contains detailed proofs of lemmas.


\section{Problem Statement}
We now define the problem and its notation, which are illustrated in Figure 1. Consider a planar rigid body of mass $m$ and radius of gyration $\rho$, so that its moment of inertia about the center of mass is $I_c= m \rho^2$. The body is supported by two unilateral frictional contact points $\vp_1,\vp_2 \in \real^2$ lying along two straight stationary segments. 
The world-fixed frame is chosen such that the $x$ axis is aligned with the line $\vp_1 \vp_2$.
At the reference configuration, the body's center of mass $\vr_c$ is located at height $h$ above the contacts, and horizontal distances from $\vr_c$ to the contact $\vp_i$ are denoted by $l_i$ for $i=1,2$, as shown in Figure 1(a). The unit vector tangent to the $i^{th}$ contact is denoted by $\vt_i$ and the normal vector is $\vn_i$, such that $\vt_i$ makes an angle of $\phi_i$ with the $x$ axis, for $i=1,2$. External forces and torques are acting on the body, and their net effect is summed by a force $\vf_{ex}$ and torque $\tau_{ex}$ about the center of mass. The force $\vf_{ex}$ makes an angle of $\Ag$ with $-z$ axis. Reaction forces $\vf_1$ and $\vf_2$ are acting on the body at the contact points. The contact forces are subject to Coulomb's dry friction law \cite{lotstedt}, and thus they must satisfy the inequality condition \beq{ineq.fric} |\vf_i \cdot \vt_i| \leq \mu_i (\vf_i \cdot \vn_i) \mbox{, for } i=1,2 \eeq where $\mu_i$ is the coefficient of friction at the $i^{th}$ contact. Geometrically, this means that the direction of contact forces $\vf_i$ are constrained to lie within the {\em friction cones} centered about the normals $\vn_i$ with half-angles of $\tan^{-1}(\mu_i)$, as illustrated in Figure 1a. The coordinates that describe the configuration of the body are chosen as $\vq=(x,z,\tg)^T$, where $\vr_c=(x,z)$ denote the position of the body's center of mass and $\tg$ is its orientation angle. The dynamic equations of motion of the body are given by
\beq{dyn} \left\{ \begin{array}{l} \vf_1 + \vf_2 + \vf_{ex} = m \vrdd_c \\[5pt] (\vr_c -\vr_1)^T \vJ \vf_1 + (\vr_c -\vr_2)^T \vJ \vf_2 + \tau_{ex} = m \rho^2 \tdd \; \end{array} \right. \;\;
\eeq
where
\beq{dyn2}
 \vJ = \matwo{0}{-1}{1}{0}.
\eeq
In order to obtain the motion of the contact points, the following kinematic relations are used. For simplicity it is assumed that contact is made only at ``vertex points'' $\vr_1,\vr_2$ which are body-fixed points of the body (e.g. endpoints of rigid ``legs'' as illustrated in Figure 1). The positions of these vertex points at the reference configuration $\vq \seq 0$ are given by $\vp_1$ and $\vp_2$, and their motion depends on the body's translation and rotation according to the kinematic relation
\beq{kin}
\vr_i(\vq) = \vr_c + \vR(\tg)\vp_i, \mbox{ for } i=1,2
\eeq
where
\beq{kin2}
\vR(\tg) = \matwo{\cos \tg}{-\sin \tg}{\sin \tg}{cos \tg}.
\eeq
Differentiating \eq{kin} twice with respect to time and using the dynamic equations \eq{dyn}, the accelerations of the two vertex points are obtained as
\beq{kin_a}
\begin{array}{lll}
\vrdd_i & = & \vrdd_c + \vRdd(\tg)\vp_i \\
&=& \vrdd_c + \left(
\tdd \vJ \vR(\tg) - \td^2 \vR(\tg)
\right)\vp_i
\\[5pt]
& = & \frac{1}{m}(\vf_1 + \vf_2 + \vf_{ex}) +
 \left[
 \frac{1}{m \rho^2}
 \left(-(\vR(\tg)\vr_1)^T \vJ  \vf_1
 \right.\right.-...\\
 & & \left.\left. (\vR(\tg)\vr_2)^T \vJ \vf_2 + \tau_{ex}\right)
 \vJ \vR(\tg) - \td^2 \vR(\tg) \right]\vp_i
\end{array}
\eeq
The tangential and normal displacements of the contacts are defined as: 
\beq{xi} x_i(\vq) \seq (\vr_i(\vq)-\vp_i) \cdot \vt_i \eeq
\beq{zi}
 z_i(\vq) \seq (\vr_i(\vq)-\vp_i) \cdot \vn_i
\eeq
for $i=1,2$, (see Figure 1b).

The relations between contact forces and contact displacements can be described by linear complementarity formulation (cf. \cite{anitescu97,blumentals2016,trinkle.comp}) as:
\beq{compl}
\begin{array}{l}
0 \leq (\vf_i \cdot \vn_i) \perp z_i \geq 0 \\
0 \leq (\vf_i \cdot \vn_i) \perp \zd_i \geq 0 \mbox{, if } z_i=0\\
\vf_i \cdot \vt_i = - \mu_i {\rm sgn}(\xd_i) (\vf_i \cdot \vn_i)
\end{array}
\eeq
for $i=1,2$, where the sign function is set-valued at zero: $\rm{sgn}(0) \in [-1,1]$. 
The state space of all positions and velocities $(\vq,\vqd)$ is bounded by the kinematic constraints of contact, and the {\em contact-feasible} space is defined as
\beq{noname1}
\begin{array}{lll}
\Fs &=& \left\{(\vq,\vqd) \in \real^{6}: \; z_i(\vq)\geq 0 \mbox{ and }\right....\\
 &&\left.\zd_i(\vq,\vqd)\geq 0\mbox{ if }  z_i(\vq) = 0 \mbox{, for } i=1,2. \right\}
\end{array}
\eeq

The configuration $\vq \seq  0$ is called an {\em equilibrium point} if there exist contact forces $\vf_1,\vf_2$ satisfying the frictional inequalities $\eq{ineq.fric}$ that also satisfy the equations \eq{dyn} under zero accelerations $\vrdd_c=0$ and $\tdd =0$. Note that the contact forces at equilibrium are non-unique, since \eq{dyn} implies a system of $3$ scalar equalities in $4$ unknowns (i.e. statical indeterminacy). Moreover, typically there exists a one-dimensional set of feasible two-contact equilibrium configurations 
described by the constraints $z_1(\vq)\seq z_2(\vq)\seq0$. That is, $\vq \seq 0$ is usually not an isolated equilibrium point.

\begin{table*}
\centering{
\begin{tabular}{|c|l|l|l|l|}\hline
   Letter & contact mode & kinematic & equalities & consistency
   \\
   & & admissibility & & constraints \\   \hline
  {\bf S} & sticking & $z_i = 0$ & $\zdd_i = 0$ & $|\vf_i \cdot \vt_i| < \mu_i (\vf_i \cdot \vn_i)$ \\ & &  $\zd_i \seq \xd_i \seq 0$ & $\xdd_i=0$ &  \\ \hline
{\bf F} & free & $z_i \geq 0$ and & $\vf_i = 0$ & $\zdd_i > 0$ if $z_i \seq \zd_i \seq 0$\\
& &$\zd_i \geq 0$ if $z_i \seq 0$ & &  \\ \hline
{\bf P} & positive slip & $z_i\seq \zd_i \seq 0$, & $\zdd_i =0$, & $\vf_i \cdot \vn_i > 0 $,\\ & & $\xd_i \geq 0$ & $\vf_i \cdot \vt_i =  -\mu (\vf_i \cdot \vn_i)  $& $\xdd_i > 0$ if $\xd_i \seq 0$\\ \hline
{\bf N} & negative slip & $z_i\seq \zd_i \seq 0$, & $\zdd_i =0$, & $\vf_i \cdot \vn_i > 0 $,\\ & & $\xd_i \leq 0$ & $\vf_i \cdot \vt_i =  \mu (\vf_i \cdot \vn_i)  $& $\xdd_i < 0$ if $\xd_i \seq 0$\\ \hline
\end{tabular}
\caption{table of contact modes for a single contact}}
\label{tab.modes}
\end{table*}

In order to analyze the behavior of solution trajectories near equilibrium, one has to define a {\em distance metric} $\Delta(\vq,\vqd)$ that measures the distance of a state $\vq(t), \vqd(t)$ from the equilibrium state $\vq \seq \vqd \seq 0$. The distance $\Delta$ can be chosen, for instance, as the Euclidean norm in $\real^6$, but some other valid choices also exist which do not necessarily satisfy the properties of a norm. Any choice of a distance metric $\Delta$ enables one to introduce the notion of finite-time Lyapunov stability (FTLS) of an equilibrium configuration, which is defined as follows:
\begin{defn}
Let $\vq \seq 0$ be an equilibrium configuration of a planar rigid body on two frictional contacts. This configuration is called \textbf{finite-time Lyapunov stable (FTLS)} if for every arbitrarily small $\epsilon > 0$ there exists $\delta>0$ such that for any initial position-and-velocity perturbation $(\vq(0),\vqd(0)) \in \Fs$ that satisfies \\
$\Delta(\vq(0),\vqd(0))<\delta $,
 the solution $\vq(t),\vqd(t)$ of \eq{dyn} satisfies $\Delta(\vq(t),\vqd(t))<\varepsilon $ for all $t>0$. Moreover, the solution must reach a static equilibrium configuration where $\vqd \seq 0$ in a finite time $t_f$ that satisfies $t_f < \varepsilon$.
\end{defn}
The definition of FTLS is very similar to the classical notion of Lyapunov stability of equilibria in dynamical systems theory \cite{lyapunov_book}, with an additional requirement of finite-time convergence to an equilibrium state in the vicinity of the original configuration. Despite the simplicity of FTLS definition, the dynamics in \eq{dyn} turns out to be highly complicated. The unilateral contacts and friction constraints \eq{ineq.fric} make this a hybrid dynamical system which undergoes state transitions between different modes of contacts. Moreover, the velocities $\vqd(t)$ are also piecewise continuous due to the occurrence of collisional impacts at the contacts. In the next three sections we explicitly formulate the hybrid dynamics of this two-contact rigid body system, including contact mode transitions and impacts at the contacts.

\section{Contact dynamics, ambiguous equilibria and instability}

We now explicitly formulate the dynamics of the system under all possible contact modes, 
as implied by the complementarity relations \eq{compl}. 
Then we demonstrate possible ambiguity of static equilibrium solutions and prove its relation to instability. Finally, we define the zero-order approximation of the dynamics in a small neighborhood of the equilibrium state $\vq \seq \vqd \seq 0$. We begin by introducing all possible contact modes.

\subsection{Contact modes}
Each contact can have four different modes, denoted by $\{ {\rm F,S,P,N} \}$, which are: free, sticking, positive slip and negative slip, respectively. Each contact mode involves kinematic constraints on the contact position, velocity and acceleration, and additional constraints of the components of the contact force $\vf_i$, as summarized in Table 1. Each contact mode for two contacts is represented by a two-letter word from the alphabet $\{ {\rm F,S,P,N} \}$. For example, contact mode PF means that the contact $\vr_1$ slips forward while the contact $\vr_2$ is free. The contact mode SS corresponds to static equilibrium. Importantly, not all 15 remaining combinations of non-static contact modes are kinematically feasible in the vicinity of the $\vq = 0$ configuration. The contact modes $\{\mathrm{SP,SN,PS,NS}\}$ are associated with kinematic constraints which are generically over-constraining. From the contact modes related to simultaneous slippage on two contacts, the contact modes $\{\mathrm{PN,NP}\}$ are kinematically infeasible if $\cos \phi_1 \cos \phi_2 > 0$, or alternatively the contact modes $\{\mathrm{PP,NN}\}$ are kinematically infeasible if $\cos \phi_1 \cos \phi_2 < 0$. That is, only nine non-static contact modes are kinematically feasible on generic geometric arrangements of the two contacts. Unlike the static contact mode SS for which the contact forces $\vf_i$ are indeterminate, for each choice of non-static contact mode a unique solution for the body's accelerations $(\vrdd_c,\tdd)$ and contact forces $\vf_i$ is obtained according to the following procedure. For given positions $\vq$ and velocities $\vqd$ of the body, one first has to verify that the chosen contact mode is kinematically admissible according to the equalities and inequalities in column 3 of Table 1. Next, each contact mode adds two equality constraints per contact according to column 4 of the table. Substituting these equalities into the kinematic relations \eq{kin_a} and combining with the dynamic equations of motion \eq{dyn}, one obtains a linear system of 7 equations in the 7 scalar unknowns $(\vrdd_c,\tdd,\vf_1,\vf_2)$, from which a unique solution can be generically obtained. However, for any admissible contact mode the solution must also be checked for consistency according to the inequalities that appear in column 5 of Table 1, and contact modes with inconsistent solutions are excluded. This procedure inspires the notion of consistent modes:
\begin{defn}\label{def:consistency}
A given  contact mode in a given state of a system is called {\bf consistent} if the associated kinematic admissibility conditions are satisfied and there is a unique pair of accelerations and contact forces (non-static contact modes) or infinitely many of them (SS mode) satisfying the equality constraints and the consistency conditions of the contact mode.
\end{defn}
For example, an \emph{equilibrium point} as defined in Sec. 2 is equivalent to a \emph{state where the SS mode is consistent}.

A well-known observation is that the dynamics under unilateral frictional contacts may lead to peculiar cases of indeterminacy where solutions under different non-static contact modes are simultaneously consistent, or inconsistency where no contact mode generates a consistent solution. These scenarios are related to the paradox of \Painleve \cite{genot99,leine2002,mason&wang}. Conditions for  occurrence of this paradox have been analyzed in previous works \cite{or.RCD2014,or&rimon.NODY2012}. 
In particular, it has been proven in \cite{or.RCD2014} that \Painleve paradox associated with slippage on a single contact $\vp_i$ is avoided if the friction coefficient satisfies the upper bound
\beq{bound.painleve}
\mu_i<\dfrac{\kg_i^2 +\sin^2 \tg_i}{|\sin \tg_i \cos \tg_i|}
\eeq
where $\kg_i \seq \rho / ||\vr_c-\vp_i||$, $\rho$ is the body's radius of gyration, and $\tg_i$ is the angle between the contact normal $\vn_i$ and the vector $\vr_c-\vp_i$. For a uniform slender rod, \eq{bound.painleve} implies that \Painleves paradox can occur only if  $\mu \geq 4/3$, which is unrealistically large friction \cite{genot99,leine2002}. Conditions for avoiding the scenario of \Painleve paradox associated with simultaneous slippage at two contacts are more complicated \cite{blumentals2016}. Nevertheless, if the system satisfies the conditions for persistent equilibrium defined later in Section 5, then it follows from the results of \cite{varkonyi.painleve2c.2016} that this scenario can always be ruled out. Therefore, none of the situations where the solution reaches any paradox of indeterminacy or inconsistency with nonzero velocities is considered in our analysis, for the sake of simplicity.

%


\subsection{Ambiguous equilibria and instability}

Consider a two-contact equilibrium state $\vq \seq \vqd \seq 0$ for which the static contact mode SS is consistent. At this configuration, the kinematic admissibility constraints (column 3 in Table 1) are satisfied for all non-static contact modes. Therefore, the consistency of each non-static contact mode is determined by inequalities on \emph{contact forces} $\vf_i$ and on \emph{accelerations} at the contacts $\xdd_i$ and/or $\zdd_i$ (column 5 of Table 1).
This may gives rise to an important form of non-uniqueness:
\begin{defn} \label{def:ambiguity} An {\bf ambiguous equilibrium} is a static state ($\vqd=0$) in which the SS mode and one or more other contact modes are simultaneously consistent.
\end{defn}
Note that these situations are 
different from the classical \Painleve paradox mentioned above, which involves nonzero velocities.


\noindent {\bf Example 1 - ambiguity:}
Figure 2(a) shows an illustration of a sitting human who carries a heavy backpack and supports himself by contacts at the seat and on the ground%
\footnote{drawing is courtesy of Frits Ahlefeldt, \url{http://hikingartist.com}}. When the center of mass of human+backpack goes backwards beyond the support on the seat, the human can tip over while the front ground-foot contact is detaching (contact modes SF, see arrows). However, if friction is sufficiently large, a static equilibrium solution may also be consistent. This scenario is demonstrated in the two-contact configuration in Figure 2(b) for friction coefficient of $\mu_1 \seq \mu_2 \seq 0.5$. The radius of the circle represents the radius of gyration of the human$+$backpack, and the circle's center denotes the center of mass position, which is located beyond the upper contact point. Gravity force acts at the center of mass with zero torque. 
The two arrows emanating from the contact points denote contact reaction forces that balance the external load while satisfying the friction constraints \eq{ineq.fric}, hence the contact mode SS of static equilibrium is consistent. On the other hand, the dashed arrow denotes the contact force under contact mode SF of tipover motion, which is also consistent. A similar example that demonstrates ambiguity of static equilibrium with slipping contact modes NF and PF can be found in \cite{or&rimon.icra08a}.

\begin{figure}[t]
\centering{\includegraphics[width=0.47\textwidth]{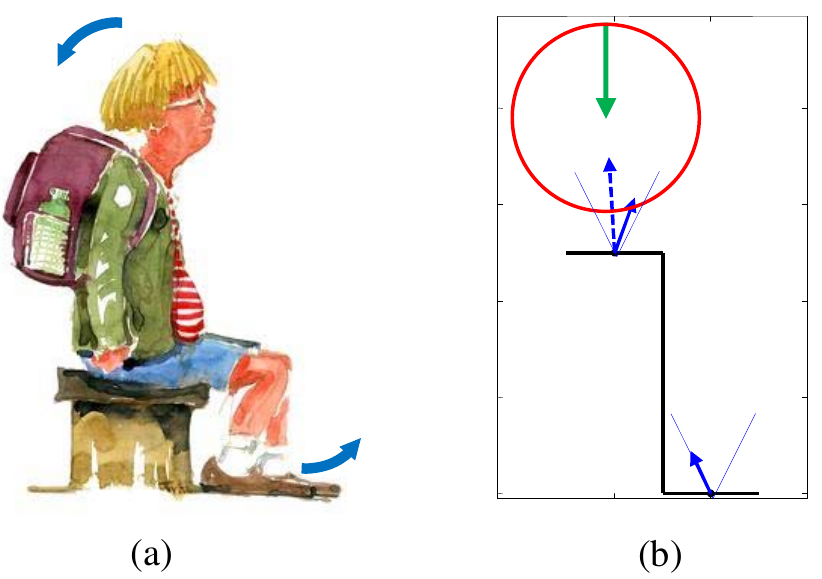}
\caption{Ambiguous equilibrium with two frictional - the heavy backpack example 1: (a) Illustration. (b) Contact sketch with SS-SF ambiguity}} \label{fig.ambig}
\end{figure}

We now review a key result which has already been presented in \cite{or&rimon.icra08a}: ambiguous equilibrium directly implies instability.
\begin{thm}[\cite{or&rimon.icra08a}] \label{thm.strong} Consider a planar rigid body under two unilateral frictional contacts. If the equilibrium state $\vq \seq \vqd \seq 0$ is ambiguous with any non-static contact mode, then it does {\bf not} possess finite-time Lyapunov stability.
\end{thm}
The proof of this theorem appears in the Appendix. It is based on the observation that if a non-static contact mode is consistent at zero velocities, then by continuity of the mode's dynamic equations with respect to state variables, it is also consistent for sufficiently small nonzero velocities and the solution diverges away and cannot be bounded within a small neighborhood of the equilibrium state. The continuity argument used in this theorem indicates that the behavior of solutions near an equilibrium state can be determined by using a zero-order approximation of the dynamics, as explained next.

\subsection{Zero-order dynamics (ZOD)}

The dynamics in a small neighborhood of the equilibrium state $\vq=\vqd =0$ is closely approximated by its zero-order expressions, in which the dynamic equations \eq{dyn} and the kinematic relations of contact accelerations $\vrdd_i$ in \eq{kin_a} are evaluated at the nominal position $\vq=0$ and zero velocities $\vqd=0$. Due to continuity of the dynamics under each particular contact mode
with respect to the state variables, the approximation error can be made arbitrarily small by choosing a sufficiently small neighborhood of the equilibrium state. Thus, our analysis will use the zero-order approximation of the dynamics (ZOD) in order to investigate Lyapunov  stability which is essentially local in nature, i.e. it involves arbitrarily small neighborhoods of initial perturbations about the equilibrium state. Under the ZOD approximation, the accelerations of the contact points are obtained from \eq{kin_a} as
\beq{ZOD_a_i}
\vrdd_i = \frac{\vf_1 + \vf_2 + \vf_{ex}}{m} + \frac{-\vp_1^T \vJ \vf_1 - \vp_2^T \vJ  \vf_2 + \tau_{ex}}{m \rho^2} \vJ  \vp_i.
\eeq
Importantly, for each contact mode, evaluation of the solution at $\vq=\vqd =0$ implies that the accelerations and contact forces are not state-dependent (i.e. they are constant).

It has been pointed out in Sec. 3.2 that we focus on non-ambiguous equilibrium configurations throughout the rest of this work. This implies that all non-static contact modes are inconsistent at $\vq \seq \vqd \seq 0$. In particular, the inconsistency of contact mode FF implies that either $\zdd_1$ or $\zdd_2$ under this contact mode must be negative (see Table 1). Without loss of generality, we choose the contact indices such that $\zdd_1<0$ under FF mode. Substituting the equality constrints $\vf_1 =\vf_2 =0$ of the FF mode into \eq{ZOD_a_i}, this implies the inequality
\beq{ineq_z1dd}
\zdd_1^{FF} = \vn_1 \cdot (\rho^2 \vf_{ex} + \tau_{ex} \vJ \vp_1) <0.
\eeq
This assumption turns out to be crucial for our \Poincare map analysis in sections 5 and 6.

\section{Inelastic impacts}
In this section, we briefly introduce our model of inelastic impact at collisions.
Each collision at the $i^{th}$ contact where $z_i \seq 0$ and $\zd_i <0$ implies impulsive forces $\vfh_i$ acting along very short times at one or both contacts, whose magnitude is typically much larger than the external loads. These contact impulses cause an instantaneous jump in the velocities $\vqd$, according to the impulse-momentum relation given by
\beq{impulse}
\begin{array}{l}
m(\vrd_c^+ - \vrd_c^-) = \vfh_1 + \vfh_2 \\[5pt]
m \rho^2(\td^+ - \td^-) =  -\vp_1^T \vJ \vfh_1 -\vp_2^T \vJ \vfh_2,
\end{array}
\eeq
where the superscripts '$-$' and '$+$' denote values right before and right after the collision, respectively. Note that the impulse-momentum equation \eq{impulse} has been evaluated at $\vq \seq 0$, that is, it is also a zero-order approximation.

We assume an inelastic impact, so that $\zd_i^+ \seq 0$. The contact impulses satisfy complementarity relations which are analogous to \eq{compl}, as:
\beq{compl.impact}
\begin{array}{l}
0 \leq (\vfh_i \cdot \vn_i) \perp z_i \geq 0 \\
0 \leq (\vfh_i \cdot \vn_i) \perp \zd_i^+ \geq 0 \mbox{, if } z_i \seq 0 \\
\vfh_i \cdot \vt_i = - \mu_i {\rm sgn}(\xd^+_i) (\vfh_i \cdot \vn_i)
\end{array}
\eeq
for $i=1,2$. The second relation in \eq{compl.impact} is needed in order to account for the common scenario where a collision at one contact occurs, i.e. $z_i \seq 0,\; \zd_i^-<0$, while the other point is already in sustained contact, i.e. $z_j \seq \zd_j^- \seq 0$, where $j \seq 3-i$. The relations in \eq{compl.impact} may result in two possible types of impacts -- a single-contact collision where $\vfh_j \seq 0$ (denoted as IF or FI impact in analogy to the continuous-time contact modes), and a double-contact collision (denoted as II). Both impacts can result in either sticking ($\xd_i^+ \seq 0$) or slippage at the contacts, depending on the friction coefficients. In the case of a single-contact impact, the resulting impact law is equivalent to that of Chatterjee \cite{chatterjee&ruina_1998} under zero coefficients of normal and tangential restitution, and also close to Routh's impact law \cite{mason&wang_impact} except for the special case of slip reversal (i.e. when $\xd^-_i\xd^+_i<0$). 
Nevertheless, we follow here the complementarity formulation in Glocker and Pfeiffer \cite{glocker1995multiple} and Leine \cite{leine.NODY2008}, which also accounts for multi-contact impacts. Cases where this impact law leads to inconsistencies are typically associated with \Painleve paradox, and thus they are not considered in this work.
Nonuniqueness of the solution is however possible. In the case of multiple solutions, we set up an a priori preference list by prefering single-impact solutions (IF,IF) over two-contact impacts (II), and sticking impacts over slipping ones.

The stability analysis of this paper could be performed with many other impact laws as well. However, the law defined above has a few key properties which will lead us to invariance relations and simplify the analysis. First,
the assumption of inelasticity will allow us to find a low-dimensional \Poincare section. Second, the dimensionality of the impact map will be reduced by exploiting its property of degree-1 homogeneity in velocities, i.e. it takes the form $\vqd^+ = \vA(\vqd^- / | \vqd^-|) \vqd^-$. The matrix $\vA$ might be piecewise-constant in the direction of  $\vqd^-$ in $\real^3$. In other words, it is invariant under multiplying $\vqd^-$ by any positive scalar. Third, some properties of the \Poincare map are implied by the fact that the impulses associated with {\em slipping impacts} in a given direction ($\xd_i^+ >0$ or $\xd_i^+ < 0$) are independent of the magnitude of pre-impact tangential velocity (in analogy with Coulomb's law for sliding friction).

\section{Analysis of hybrid contact dynamics}

\begin{figure}
\centering{\includegraphics[width=0.35\textwidth]{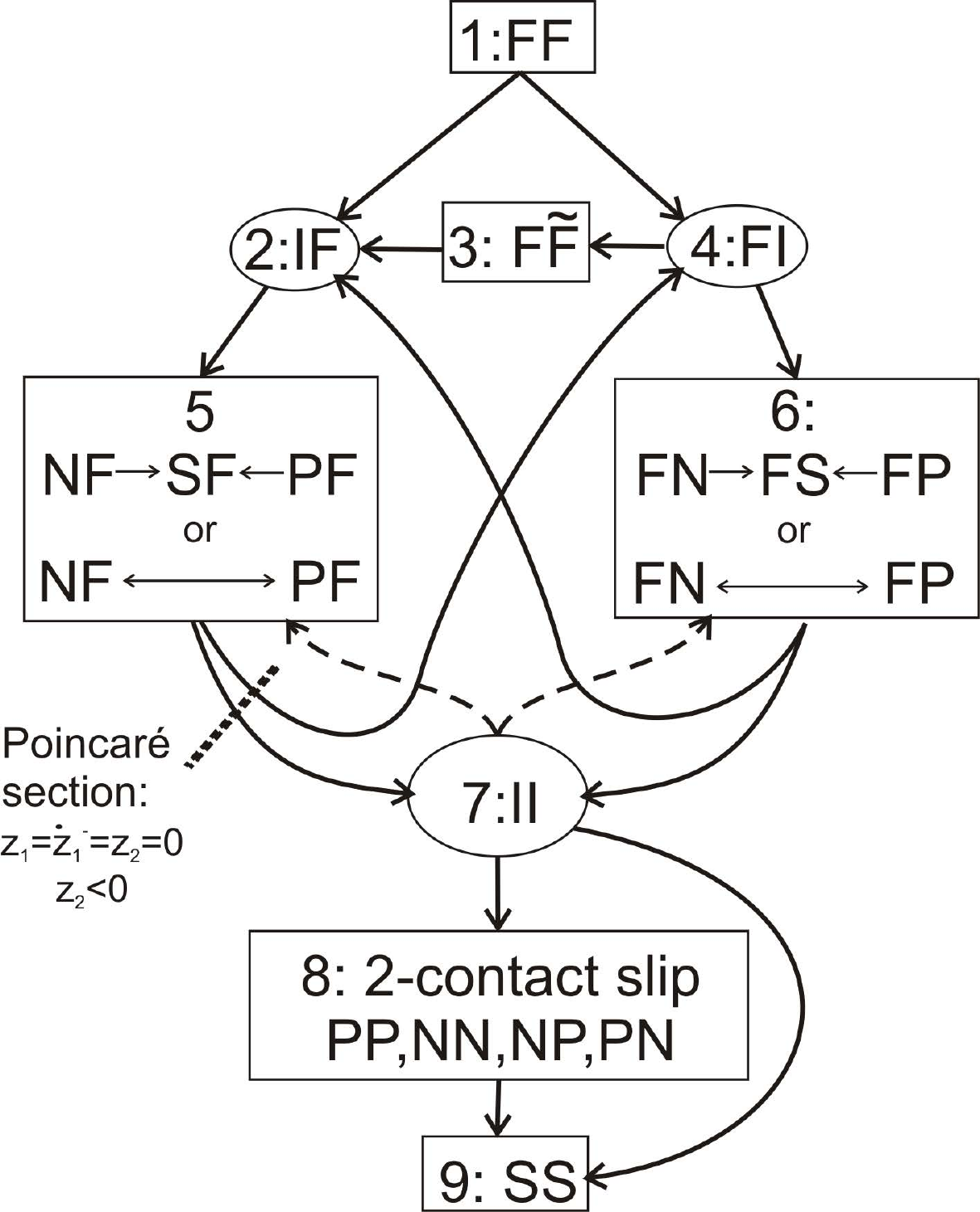}
\hspace{.1in}
\caption{Transition graph of the hybrid dynamics under zero-order approximation and assumption \eq{ineq_z1dd} of non-ambiguous equilibrium. Rounded nodes denote instantaneous impacts, and rectangular nodes represent continuous-time motion. The dashed edges are discarded in persistent equilibria, and symmetry of the graph between contacts 1 and 2 is broken by assumption \eq{ineq_z1dd}.}} \label{graph}
\end{figure}

The solution of the two-contact problem under given initial conditions undergoes transitions between different contact modes. Some of these transitions occur in continuous time (e.g. slip $\to$ stick) while others occur instantaneously via impacts. A convenient way to describe these hybrid transitions is by using a transition graph, as shown in Figure 3. Importantly, note that this transition graph is constructed under the zero-order approximation of the hybrid dynamics, and also under the assumption of non-ambiguous equilibrium which implies the inequality \eq{ineq_z1dd}. These assumptions lead to elimination of some transitions which are possible in general, and induce some asymmetry between the two contacts. Explanation on the construction of the transition graph in Figure 3 is given as follows.

Continuous-time motion under the different contact modes is denoted by rectangular nodes. These include node 1 representing the FF mode; node 5 representing all contact modes where only contact 1 is sustained  (NF, SF, PF); node 6 where only contact 2 is sustained (FN, FS, FP); node 8 representing non-static contact modes with two slipping contacts (NN, PP, NP, PN); and node 9 for the static equilibrium state SS. The last one is important as every possible motion must end there for achieving FTLS stability. Nodes 5 and 6 represent several contact modes each, among which various transitions, including slip-stick and slip reversal are possible. These possibilities are represented by arrows within the node. 
Stick $\to$ slip transition or contact detachment are not possible under the ZOD approximation for which the contact forces are constant, since these transitions are induced by varying forces. Painlev\'{e}-related singularities like dynamic jamming \cite{or.RCD2014,or&rimon.NODY2012} and ``impact without collision'' \cite{brogliato} events also require varying contact forces and thus they are impossible. Stick $\to$ slip transition is only possible in the special case where a single-contact impact results in sticking, but the contact mode SF or FS is inconsistent. This leads to immediate transition to slippage without further switches to slip reversal or sticking.

Impacts are represented by rounded nodes, and their related transitions are explained as follows. Motion in FF mode is terminated by a collision at contact 1, denoted IF (node 2) or collision at contact 2 denoted FI (node 4). Reaching a simultaneous collision at both contatcs from FF mode  is nongeneric, and thus it is not considered here. Motion with one sustained contact (nodes 5,6) must be terminated by collision of the other contact, where both contacts are in touch prior to collision. According to our impact model in the previous section, the collision can result in a two-contact impact (denoted by 'II', node 7), or a single contact impact (IF or FI, nodes 2 and 4).

Transitions occurring immediately after impacts are explained as follows. The two-contact impact II can be followed by complete stop (SS, node 9) or two-contact slippage (node 8). Another type of transition that might be possible after II impact is to motion that involves slip at one contact and separation at the other contact (NF,PF,FN, or FP), with zero initial values of the normal displacement and velocity, $z_j \seq \zd_j \seq 0$. These transitions are represented by the dashed arrows that connect node 7 back to nodes 5 and 6. Transition from II impact back to complete separation FF is ruled out by assumption \eq{ineq_z1dd} which is implied by exclusion of ambiguous equilibria. Single-contact impact FI (node 4) can be followed by motion under sustained contact at $\vr_1$ (node 6). Alternatively, a transition back to FF can be possibly made, provided that $\zdd^{FF}_2>0$ under this mode. This transition is represented by a special node (number 3) which is denoted by $F\tilde F$ in order to reflect the fact that this motion starts with the particular initial conditions of $z_2 \seq \dot z_2 \seq 0$. During this motion, contact 2 is separating while contact 1 is accelerating towards collision due to assumption \eq{ineq_z1dd} which implies $\zdd_1<0$. Therefore, motion under $F\tilde F$ must end by a single-contact impact IF (node 2). After such an impact, the only possible transition is to motion under sustained contact at $\vr_2$ (node 5), where a transition to FF mode is ruled out by using assumption \eq{ineq_z1dd}. Importantly, a similar transition after IF impact (analogous to $4 \to 3$) is impossible due to assumption \eq{ineq_z1dd}. This difference induces asymmetry between the contacts into the graph, which will be exploited in the next section. Finally, for some rare combinations of model parameters and initial conditions, the transitions $4\rightarrow 6$ and $4\rightarrow 3$ are simultaneously consistent, i.e. the system exhibits dynamical indeterminacy.  This scenario is associated with \Painleve paradox, and thus it is not considered in our analysis. 

Motivated by the transition graph, we now introduce a subclass of non-ambiguous equilibrium configurations, called {\em persistent equilibria}, which are defined as follows.

\begin{defn} \label{def:persistence}
Let $\vq \seq 0$ be an equilibrium configuration of a planar rigid body on two unilateral frictional contacts. This configuration is called {\bf persistent equilibrium} if it satisfies the following requirements:
\begin{enumerate}
  \item It is a non-ambiguous equilibrium. 
  \item Under the ZOD of each contact mode with a single slipping contact ({\rm PF,NF,FP} and {\rm FN}), either the normal force at the slipping contact satisfies $\vf_i \cdot \vn_i <0$, or the normal acceleration of the other contact (which is in F mode) satisfies $\zdd_j<0$.
  \item Under the ZOD of each contact mode with two slipping contacts which is kinematically feasible ({\rm PP $\&$ NN} or {\rm PN $\&$ NP}), the normal forces at both contacts satisfy $\vf_i \cdot \vn_i >0$ for $i=1,2$.
\end{enumerate}
\end{defn}
The implication of requirements 2 and 3 in this definition is that after a two-contact impact (II), the transitions back to nodes 5 or 6 which are represented by dashed arrows are ruled out from the transition graph in Figure 3. Moreover, these requirements also imply that once the solution reaches motion of two-contact slippage (node 8 in the graph), it stays there and slippage is decelerated until full stop at static equilibrium SS.

Under small initial perturbation from a persistent equilibrium state, the dynamic response can undergo transitions according to the transition graph in Figure 3, excluding the dashed arrows. The continuous time spent in each node is not represented in this graph, as well as divergence of the solution from the original equilibrium point. Bounds on these two quantities are given in lemma \ref{lem:finite-impacts} below. For convenience, we first define an alternative set of coordinates $\vq'=(z_1,z_2,x_2)$ for describing the motion in a small neighborhood near an equilibrium state. A transformation between the original coordinates $\vq=(x,z,\tg)$ and the new coordinates $\vq'$ always exist locally
, provided that $\cos \phi_1 \neq 0$. 
(This condition is violated only in the non-generic case where the line of contact normal at $\vp_1$ intersect the other contact point $\vp_2$). Using the new coordinates $\vq'$, we define a distance metric $\Delta$ from the equilibrium state $\vq' \seq \vqd' \seq 0$ as:
\beq{Delta}
\Delta(\vq',\vqd')=\max\left(\sqrt{z_{1}},\sqrt{z_{2}},\sqrt{|x_{2}|},|\dot{z}_{1}|,|\dot{z}_{2}|,|\dot{x}_{2}|\right).
\eeq
In addition to the metric $\Delta$, we introduce two pseudometrics, which are defined as:
\beq{dD}
\begin{array}{l}
d(\vq',\vqd')=\max\left(\sqrt{z_{1}},\sqrt{z_{2}},|\dot{z}_{1}|,|\dot{z}_{2}|\right)\\[5pt]
D(\vq',\vqd')=\max\left(d(\vq',\vqd'),|\dot{x}_{2}|\right).
\end{array}\eeq
The pseudometric $d$ in \eq{dD} measures distance from the set of states with two sustained contacts, while $D$ measures distance from the set of two-contact static equilibrium states in the vicinity of $\vq' \seq 0$. For any given state $(\vq',\vqd')$, these metrics are ordered as $d \leq D \leq \Delta$.
The use of these pseudometrics is necessary here, since one needs to establish Lyapunov stability of a specific equilibrium point which is embedded within a continuous set of equilibrium states, in contrast to attractivity of the entire set as in \cite{leine.NODY2008}. 
For a given solution $\vq'(t),\vqd'(t)$, we denote these metrics by $d(t),\; D(t)$ and $\Delta(t)$. The following lemma provides bounds on the pseudometrics $d(t)$ and $D(t)$ along solutions under small initial perturbations about equilibrium.

\begin{lem}
\label{lem:finite-impacts}
Let $\vq' \seq 0$ be a persistent equilibrium configuration of a planar rigid body on two unilateral frictional contacts. There exist finite positive scalars $k_{1}$ and $c_{1}$ such that any possible solution trajectory under the {\rm ZOD} assumption must satisfy the following bounds:\\
(i) if the system undergoes an impact at time $t_{1}$ then
\begin{equation}
D(t_{1}^{+})<k_{1}\cdot D(t_{1}^{-}) \mbox{ and }
d(t_{1}^{+})<k_{1}\cdot d(t_{1}^{-})
\label{eq:impact-Ddlimit} 
\end{equation}
(ii) if the systems undergoes no impact or contact mode transition
between times $t_{1}$and $t_{2}$, then
\begin{equation}
D(t)<k_{1}\cdot D(t_{1}) \mbox{ and } d(t)<k_{1}\cdot d(t_{1})
\, for\, all\, t_{1}<t<t_{2}. 
\label{eq:lemma-Ddlimit}
\end{equation}
(iii) in addition, if the systems is not in {\rm SS} mode at $t_{1}$, then
\begin{equation}
t_{2}-t_{1}\leq c_{1}\cdot D(t_{1}),\label{eq:lemma-tlimit for D}
\end{equation}
and if it is not in {\rm SS, PP, NN, PN}, or {\rm NP} mode then
also
\begin{equation}
t_{2}-t_{1}\leq c_{1}\cdot d(t_{1}).\label{eq:lemma-tlimit for d}
\end{equation}

\end{lem}
The proof of this lemma, which is based on linearity properties of the impact laws as well as the ZOD solution for each contact mode, appears in the Appendix. Note that \eqref{eq:lemma-tlimit for D} implies that the solution cannot stay at a single node other than mode SS for unbounded time, and that any solution with a finite path of mode transitions is bounded in state space. Additionally, it must reach SS in bounded time and then stay there forever. Importantly, the lemma does not cover the case of solution trajectories whose corresponding paths in the transition graph contain {\em infinitely} many nodes. Such infinite paths must contain a cycle, i.e. a recurring node.
A key observation which is directly implied by exclusion of the dashed transitions in the graph is that any path that contains cycles must exit node 5 once every cycle, either to a two-contact impact (II) or to a single contact impact (FI). This motivates the definition of a \Poincare section at this event as explained in the next section.

\section{Analysis and reduction of 2-contact Poincar\'{e} map}
In this section, we define a \Poincare map of the solution, construct its reduction into two scalar maps, and discuss some properties of these maps. Then we present two examples that demonstrate these properties and also show how solution trajectories of the system can be extracted from these reduction maps.

\subsection{Definition of the \Poincare map and its reduction maps}


Considering only persistent equilibrium configurations, the dashed transitions in the graph of Figure 3 were excluded. As stated above, any solution trajectory that contains cycles in the transition graph must exit node 5 once every cycle. Therefore, we define a \Poincare section $\Ss$ in the state space as \beq{Psec}
\Ss = \{(\vq',\vqd') \in \real^{6}: z_1=z_2=0,\; \zd_1^-=0 \mbox{ and } \zd_2^-<0 \}. \eeq
This \Poincare section represent the pre-impact states upon exit from node 5, where contact 1 is sustained while contact 2 is colliding. The section $\Ss$ is a three-dimensional linear (conic, to be precise) subspace, which is parametrized by pre-impact values of three variables, augmented in the vector $\vy = (x_2,\zd_2^-,\xd_2^-)$. The \Poincare map of the system is then defined as $\vP: \Ss \to \Ss$ which maps a point $\vy$ of initial conditions on the section $\Ss$ to the values at the next time that the state of the solution trajectory crosses $\Ss$. That is, the map induces a discrete-time dynamical system $\vy^{(k+1)} = \vP(\vy^{(k)})$ of the pre-impact states once every step of this impact event. Note that $\vP$ may be undefined for some initial conditions $\vy \in \Ss$, where the solution has a finite non-cyclic transition path that does not intersect $\Ss$ again. Lemma 1 in the previous section implies that such solutions must terminate at static equilibrium SS in a finite bounded time.

The fact that the \Poincare map $\vP$ is based on solutions of continuous-time motion and impacts under the ZOD approximation  implies two important invariance properties, which are expressed as follows:

\noindent
\textbf{Invariance with respect to $x_2$:} for any $\beta \in \real$ and $(x_2,\zd_2^-,\xd_2^-)$, the \Poincare map $\vP$ satisfies
\beq{inv_x2} \vP(\beta + x_2 ,\zd_2^-,\xd_2^-) = (\beta,0,0)+\vP(x_2,\zd_2^-,\xd_2^-). \eeq
\textbf{Scaling invariance:}
if arbitrary initial conditions of the system at $t=0$ are upscaled in such a way that all velocity coordinates are multiplied by a factor of $\beta$ and all position coordinates are multiplied by $\beta^2$, then the original trajectory $\vq(t)$ and the modified trajectory $\vq^*(t)$ will be related as $\vq^*(\beta t)=\beta^2\vq(t)$ for all $t$. This relation holds
because under the ZOD, all equations of motion are differential equations with piecewise constant right-hand sides, all impact maps are linear homogenous in velocities (see Sec. 4) and all switching surfaces (between stick and slip) and contact surfaces are given by linear homogenous functions of the state variables.
Consequently, for any $\beta >0$ and $(x_2,\zd_2^-,\xd_2^-)$, the \Poincare map $\vP$ satisfies \beq{inv_scale} \vP(\beta^2 x_2 ,\beta\zd_2^-,\beta\xd_2^-) =  \vP(x_2,\zd_2^-,\xd_2^-) \cdot \left(
\begin{array}{ccc}
\beta^2 & 0 &0 \\ 0 & \beta & 0 \\ 0 & 0& \beta \end{array}
\right). \eeq

Under these invariance properties, parametrization of the \Poincare section $\Ss$ can be reduced to a single scaled variable, defined as   \beq{varphi} \varphi = \tan^{-1}\left(\frac{\xd_2^-}{|\zd_2^-|}\right). \eeq
Physically, $\varphi$ is the angle of pre-collision velocity $\vrd_2^-$ with respect to the normal $\vn_2$, and satisfies $\varphi \in I$ where $I=(-\tfrac{\pi}{2},\tfrac{\pi}{2})$. Using the scaled variable $\varphi$, the {\em reduced \Poincare map}
$R: I \to I$ is defined as \beq{R} R(\varphi^{(k)}) = \varphi^{(k+1)}. \eeq
Another important scalar function is the {\em growth map} $G: I \to \real_+$,
defined as \beq{G} G(\varphi^{(k)}) = \frac{|\zd_2^{(k+1)}|}{|\zd_2^{(k)}|}. \eeq

The reduced \Poincare map $R(\varphi)$ and the growth map $G(\varphi)$ are scalar functions which can be plotted and visualized. Together, they encode most of the information on solution trajectories of the hybrid dynamics, as demonstrated in the sequel. Importantly, these functions are {\em semi-analytic}, since for each given value of $\vfi$, the maps are associated with a finite sequence of contact modes and impacts under ZOD approximation, and the solution can be obtained in closed form as a concatenation of constant-acceleration solutions. Moreover, each sequence of transitions can also be accompanied with closed-form inequalities that give conditions for its validity. In practice, these functions are computed numerically due to the high complexity of all possible contact transitions.

\subsection{Properties of the reduction maps}
We now discuss some important properties of the reduction maps $R$ and $G$. First, it is clear that $R$ and $G$
may be undefined for some portions of their domain $I$. This is because there exist values of $\vfi$ corresponding to initial conditions on the \Poincare section $\Ss$ which result in solution trajectories that reach the mode SS in finite time via a double impact (II) and do not cross $\Ss$ again.

Second, $R$ and $G$ can attain constant values along some sub-intervals of $I$. For example, consider the case where a cyclic path in the transition graph goes through a state where the velocities of one contact satisfy \linebreak $\xd_i \seq \zd_i \seq 0$ due to sticking.
This constraint uniquely determines the direction $\vqd' / |\vqd'|$ of the velocity vector in $\real^3$, while only its magnitude $|\vqd'|$ may vary freely. Due to linearity of the governing equations, the direction of velocities becomes uniquely determined for the rest of the motion until returning to the \Poincare section. Thus, the value of the map $R$ becomes constant for all values of $\vfi$ for which this particular transition path holds.
Sub-intervals where the growth map $G$ is constant are explained as follows. Under the ZOD, the contact force in positive or negative slip mode is independent of $x_{2}$ and $\dot{x}_{2}$. Similarly, the contact impulse of a positive or negative slipping impact does not depend on the tangential velocity of the contact point
nor does the time at which the impact occurs.
Then, if for some nominal initial value $(x_{2},\dot{z}_{2}^{-},\dot{x}_{2}^{-})$ on the \Poincare section, the cyclic path in the transition graph contains only slipping contact modes and slipping impacts without slip reversal (i.e. no transitions such as NF$\longleftrightarrow$PF or FN$\longleftrightarrow$FP), then these invariance properties imply the invariance relations:
\beq{G-endpoints-sliding} \vP(x_{2},\dot{z}_{2}^{-},\dot{x}_{2}^{-}+\beta)=\vP(x_{2},\dot{z}_{2}^{-},\dot{x}_{2}^{-})+(\beta T,0,\beta) \eeq
for a finite range of small values of $\beta$ for which the cyclic path of mode transitions remains unchanged. Here, $T$ is the time duration of the nominal cycle. Therefore, one obtains that the growth map $G$ is constant for all values of $\vfi$ for which this particular transition path holds.

The third property of $R$ and $G$ is continuity:
\begin{lem}
\label{lem:continuity_RG} the maps $R(\vfi)$, $G(\vfi)$ are continuous and piecewise smooth.
\end{lem}
The proof of Lemma \ref{lem:continuity_RG} in the Appendix is based on ruling out two possible scenarios:
\begin{enumerate}
\item discontinuity of \eq{varphi} when $\zd_2^{(k+1)}=0$
\item discontinuity of the full \Poincare map $\vP$
\end{enumerate}

The fourth property of the maps $R$ and $G$ is that they display a special behavior near the endpoints $\vfi \to \pm \pi/2$. This is summarized in the following lemma, whose proof appears in the Appendix:

\begin{lem}
\label{lem:properties_RG} If the maps $R(\vfi)$, $G(\vfi)$ are defined at an endpoint $\vfi \to \pm \pi/2$,
then there exists a finite-sized subinterval $(-\pi/2,\vfi_0]$ or $[\vfi_0,\pi/2)$ for which the growth map $G(\vfi)$ attains a constant value of
\beq{G.pi2}
G(\vfi)=G^\pm
\eeq
furthermore $R(\vfi)$ satisfies the following relations:

\beq{R.pi2} \lim_{\vfi \to \pm \tfrac{\pi}{2}}R(\vfi) = \pm \tfrac{\pi}{2} \eeq
\beq{dR.pi2} \lim_{\vfi \to \pm\tfrac{\pi}{2}}R'(\vfi) =  G^{\pm}, \mbox{  where $R'(\vfi)=dR/d \vfi$.}\eeq
\end{lem}

\begin{figure*}
\centering{\includegraphics[width=0.9\textwidth]{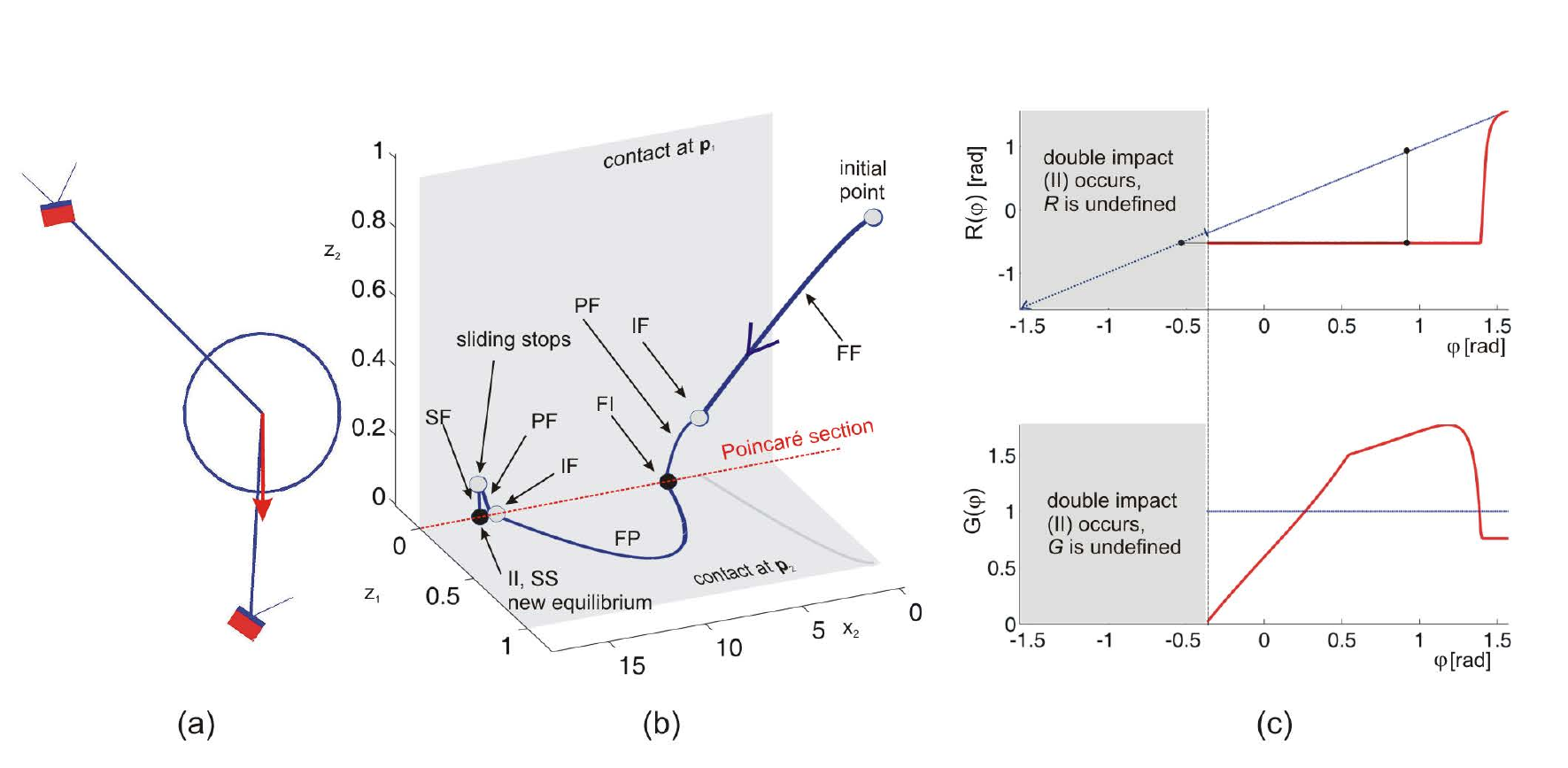}
\caption{Example 2 - (a) Two-contact equilibrium configuration. (b) A solution trajectory of $\vq'(t)$. (c) Plots of the reduced \Poincare map $R(\vfi)$ (top) and the growth map $G(\vfi)$ (bottom). }} \label{fig.example1}
\end{figure*}

Finally, a key observation regarding the reduced \linebreak \Poincare map $R(\vfi)$ is the interpretation of {\bf fixed points} which satisfy $\varphi^* = R(\varphi^*)$. These fixed points correspond to periodic solutions of the reduced discrete-time dynamics of $\varphi^{(k)}$. A known fact (cf. \cite{morris.tac2009}) is that local convergence or divergence of the series $\vfi^{(k)}$ near a fixed point $\varphi^*$ as $k \to \infty$ can be determined by checking the derivative $R'(\varphi)$ at $\varphi \seq \varphi^*$, such that the fixed point is locally convergent if $|R'(\varphi^*)|<1$, while divergence is implied by $|R'(\varphi^*)|>1$. Using Lemma \ref{lem:properties_RG} then implies that if $R$ is defined at an endpoint $\vfi \to \pm \tfrac{\pi}{2}$, then it is also a limiting fixed point of the discrete-time dynamics, whose convergence or divergence is determined by the condition $G^{\pm} <1$. Note that discrete-time convergence to $\vfi \seq \pm \tfrac{\pi}{2}$ is one-sided and attained only asymptotically, since $\vfi^{(k)}$ must always lie within the interval $I$. Importantly, fixed points of $R(\vfi)$ and their convergence only give information about behavior of the reduced discrete-time solution $\vfi^{(k)}$, and do not necessarily imply stability, as the magnitude of the full state vector at the \Poincare section $\vy^{(k)}$ may grow unbounded. Complete information on the behavior of $\vy^{(k)}$ can be extracted from the growth map $G$, by using the symmetry relations \eq{inv_x2}, \eq{inv_scale} under the ZOD assumption. This concept is demonstrated in the following examples.

\subsection{Examples}
We now present two examples of 2-contact frictional equilibrium configurations and show the corresponding plots of the reduced \Poincare map $R(\vfi)$ and the growth map $G(\vfi)$. Then we discuss the behavior of solutions by showing representative trajectories of $\vq'(t)$ and $\vfi^{(k)}$. The parameters of the contacts' geometry and external force are given using the notation of Figure 1(a). Distances are scaled by the body's radius of gyration (this is equivalent to assuming $\rho \seq 1$), which is also shown as the circle's radius. In all the examples, the external force is applied at the center of mass without additional torque ($\tau_{ex} \seq 0$) and the plots' axes are rotated by $-\Ag$ so that the external force is pointing downward, similar to gravity. Mass and force are scaled such that $m \seq 1$ and $||\vf_{ex}|| \seq 1$. The contact supports are drawn as thick lines aligned with tangential directions $\vt_i$, and the edges of the friction cones appear in thin lines.

\noindent {\bf Example 2:} the contact configuration is shown in Figure 4(a). The data of the contacts and external force is given by $\Ag \seq 65^\circ$, $l_1 \seq -3.5$, $l_2 \seq 2.25$, $h \seq 1.25$, $\phi_1 \seq 75^\circ$, $\phi_2 \seq 30^\circ$, $\mu_1 \seq 0.75$ and $\mu_2 \seq 0.6$. 
According to \eq{bound.painleve}, the conditions for avoiding \Painleve paradox are $\mu_1 < 0.846$ and $\mu_2<0.942$, which are both satisfied. 
An example of a solution trajectory in the 3D space of $\vq'=(z_1,z_2,x_2)$ under the ZOD assumption is shown in Figure 4(b). 
The initial conditions are given by $\vq'(0)=(1,1,0)$ and $\dot\vq'(0)=(0,0,2)$, which correspond to starting at the contact mode FF. The finite sequence of contact modes along the solution trajectory shown in Figure 4(b) is FF $\to$ IF $\to$ PF $\to$ FI 
$\to$ FP $\to$ IF $\to$ PF $\to$ SF $\to$ II $\to$ SS. The solution stops at a nearby static equilibrium configuration of $\vq'=(0,0,14.87)$ and $\vqd'=0$ in finite time, as implied by Lemma 1. Circles along the trajectory denote contact mode transitions, while filled circles denote points where the trajectory crosses the \Poincare section at $z_1 \seq z_2 \seq 0$, which is denoted by a dashed line. The plots of the two maps $R(\vfi)$ (top) and $G(\vfi)$ (bottom) are shown in Figure 4(c). The black circles on the plot of $R(\vfi)$ denote the discrete series $\vfi^{(1)}=0.948$ and $\vfi^{(2)}=-0.524$ of values at events where the solution $\vq'(t)$ crosses the \Poincare section. The graph of $R(\vfi)$ indicates that $R(\vfi^{(1)})=\vfi^{(2)}$ while $R(\vfi^{(2)})$ is undefined, which implies a finite trajectory which does not cross the \Poincare section again. More generally, it can be seen that both $R$ and $G$ are undefined for the range $-\pi/2 < \vfi < -0.38$. This is because initial conditions of $\vfi$ in this range result in a finite sequence of modes which ends at a two-contact impact (II) that either stops immediately at SS or stops after a finite time of two-contact slipping mode NN, so that the \Poincare section is never reached again. In the range $-0.38 < \vfi < 0.54$, the contact mode sequence is FI  $\to$ FS $\to$ FP $\to$ IF $\to$ SF $\to$ II, and in the range of $0.54 < \vfi < 1.39$ the contact mode sequence changes to FI $\to$ FP $\to$ IF $\to$ SF $\to$ II. That is, the first FI impact changes from sticking to slipping impact. This implies that $G(\vfi)$ is nonsmooth at $\vfi \seq 0.54$ due to this transition. Nevertheless, since both contact sequences contain a mode of sticking contact (SF), the function $R(\vfi)$ is constant along the entire interval $(-0.38,1.39)$ while $G(\vfi)$ is varying, as explained above. On the other hand, for $\vfi \! \in \! (1.39,\pi/2)$ the contact mode sequence changes again to FI  $\to$ FP $\to$ IF $\to$ PF $\to$ FI. Since this sequence contains only slippage without reversal, it corresponds to constant $G$ while $R(\vfi)$ is varying, as explained above.

\begin{figure*}
\centering{\includegraphics[width=\textwidth]{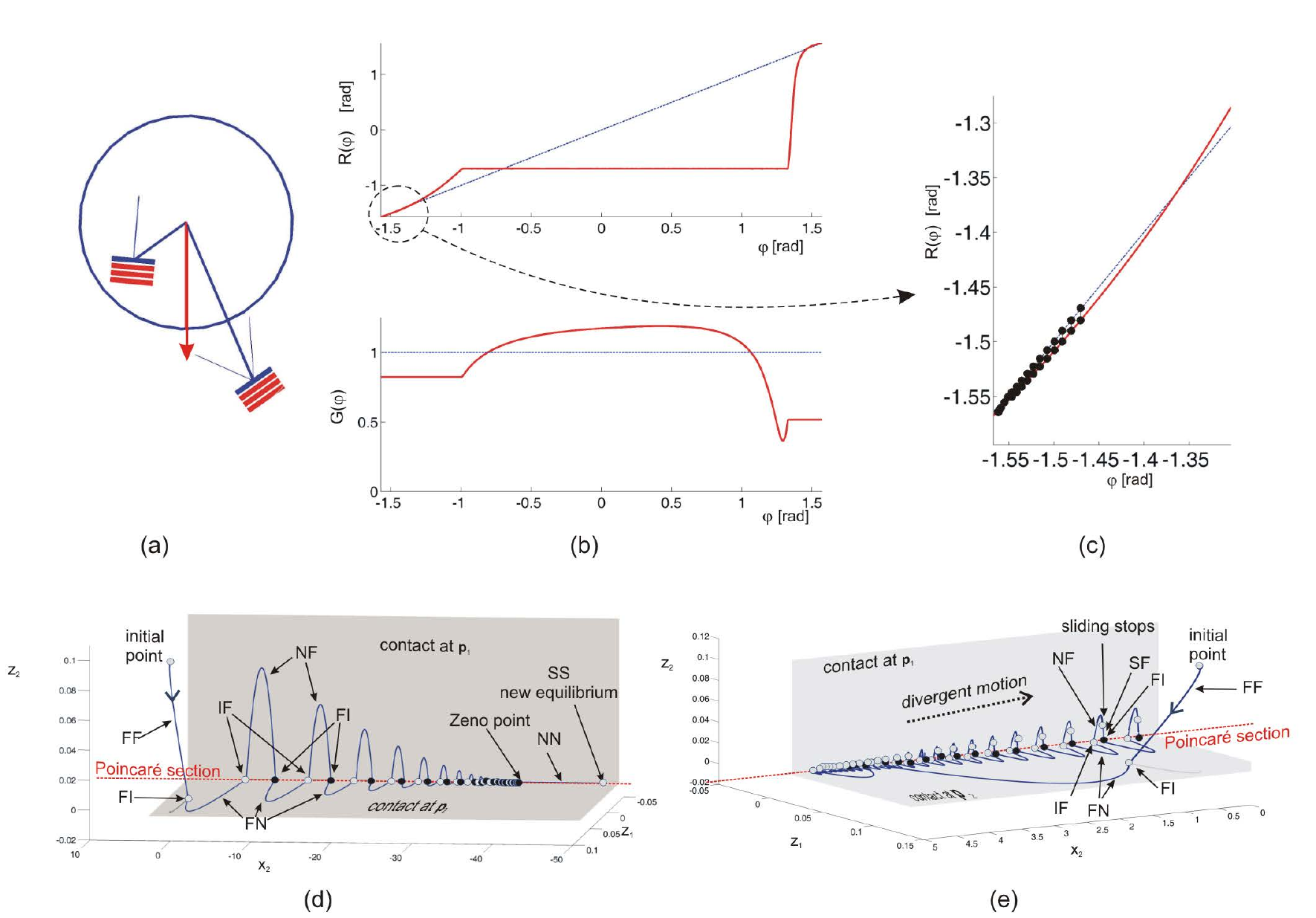}
\caption{Example 3 - (a) Two-contact equilibrium configuration. (b) Plots of the reduced \Poincare map $R(\vfi)$ (top) and  the growth map $G(\vfi)$ (bottom). (c) A converging Zeno solution trajectory of $\vq'(t)$ where $\vfi^{(k)} \to -\tfrac{\pi}{2}$, which reaches SS in finite time. (d) A solution with diverging Zeno behavior around $\vfi \seq -0.7$.}} \label{fig.example2}
\end{figure*}

\noindent {\bf Example 3:} an equilibrium configuration on two contacts is shown in Figure 5(a). The data of the contacts and external force is given by $\Ag \seq 45^\circ$, $l_1 \seq 1.5$, $l_2 \seq -0.1$, $h=0.6$ $\phi_1 \seq 80^\circ$, $\phi_2 \seq 40^\circ$, $\mu_1 \seq 0.6$ and $\mu_2 \seq 0.001$. 
According to \eq{bound.painleve}, the conditions for avoiding \Painleve \linebreak paradox are $\mu_1 < 2.12$ and $\mu_2<6.64$, which are both satisfied. 
Plots of the two maps $R(\vfi)$ (top) and $G(\vfi)$ (bottom), are shown in Figure 5(b). The maps are defined over the entire interval $I$ here. The graph of $R(\vfi)$ crosses the dashed line $R(\vfi)=\vfi$ at several points, which are fixed points of $R$. According to eq. \eq{R.pi2} in Lemma \ref{lem:properties_RG}, the endpoints at $\vfi = \pm \tfrac{\pi}{2}$ are also (limiting) fixed points of $R$. Also, according to \eq{dR.pi2} and the values of $G$ at the endpoints which are below 1, these are two convergent fixed points. A zoom into the graph of $R(\vfi)$ at the vicinity of the left endpoint is shown in Figure 5(c). It shows that there exists another nearby fixed point at $\vfi \! \approx \! -1.36$, which is divergent, since $R'>1$ at that point. The black circles in the plot denote a series of $\vfi^{(k)}$ which converges asymptotically to $-\tfrac{\pi}{2}$. This series corresponds to the solution trajectory of $\vq'(t)$ shown in Figure 5(d), which starts at contact mode FF under initial conditions $\vq'(0)=(0.1,0.1,0)$ and $\vqd'(0)=(0.1,0,-1.5)$. The trajectory is attracted to the fixed point $\vfi \to -\tfrac{\pi}{2}$, while the magnitude of $\zd_2$ is decaying exponentially (since $G^-<1$). This is precisely a {\em Zeno solution} which converges to the \Poincare section in finite time. Nevertheless, the convergence point satisfies $\zd_1 \seq \zd_2 \seq 0$ but $\xd_2 \! \neq \! 0$, Thus, after reaching this point the solution switches to the contact mode NN of two-contact slippage, and stops at static equilibrium SS in finite time. Similar behavior also occurs near the other endpoint $\vfi \to \tfrac{\pi}{2}$ (not shown). Another fixed point of $R$ at $\vfi \seq -0.7$ can be seen in Figure 5(b). This fixed point lies within a sub-interval at which $R$ is constant so that $R' \seq 0$. This implies that $\vfi \seq -0.7$ is an attractive fixed point such that the series $\vfi^{(k)}$ reaches this value and stays constant after a finite number of discrete-time steps (rather than asymptotic convergence where $R' \! \neq \! 0$). Nevertheless, the graph of $G$ indicates that $G(-0.7) > 1$, which implies that the value of $\zd_2$ at every recurrence is diverging. This is precisely a diverging Zeno solution (also called {\em reverse chatter} in \cite{nordmark.ima2010,Varkonyi.icra2012}), as shown in the trajectory of $\vq'(t)$ in Figure 5(e) under initial conditions
$\vq'(0)=(0.1,0.1,0)$ and $\vqd'(0)=(0.1,0,1.1)$. Importantly, the existence of this diverging solution implies that the equilibrium point is {\em unstable}, in spite of the existence of different initial conditions which lead to finite-time convergence to static equilibrium as in Figure 5(d). One can see that existence of attractive fixed points of $R$ for which the value of $G$ is above $1$ implies the loss of FTLS stability  for the equilibrium point at $\vq' \seq 0$. On the other hand, in the previous example of Figure 4, the equilibrium point at $\vq' \seq 0$ possesses FTLS stability, despite of existence of regions for which $G(\vfi)>1$. These observations are formalized in the next section which gives a series of FTLS stability and instability theorems based on properties of the maps $R$ and $G$.

\section{Instability and stability conditions}

We now use the reduced \Poincare map $R$ and growth map $G$ to present the main contribution of this paper - conditions for stability and instability of two-contact persistent equilibrium configurations. First, we present conservative conditions for stability and instability based on simple properties of $R$ and $G$. Then, we present a general condition for stability which is based on more detailed analysis of the discrete-times dynamics induced by the maps $R$ and $G$. Finally, we present examples of computing and visualizing regions of stability and instability in two-dimensional parameter planes.

\subsection{Conservative conditions for stability and instability }
We now present a simple sufficient condition for instability due to reverse chatter, which is summarized in the following theorem .
\begin{thm} \label{thm.instability_simple}
If the reduced \Poincare map $R(\vfi)$ associated with the ZOD in the neighborhood of an equilibrium configuration has a fixed point $\varphi^*=R(\vfi^*) \in I$ which satisfies $G(\varphi^*)>1$, then the equilibrium configuration is not FTLS.\end{thm}

\textbf{Proof:} Suppose that the equilibrium is perturbed such that the initial condition is
$z_{1}=z_{2}=\dot{z}_{1}=x_{2}=0,\; \dot{x}_{2}=\epsilon\sin\varphi^*$,
and $\dot{z}_{2}=-\epsilon\cos\varphi^*$ where $\epsilon$ is an
arbitrarily small positive number. Then under the ZOD, the solution of $\vfi^{(k)}$ stays at the fixed point $\varphi^*$, while the magnitude of the collision velocity $\zd_2^{(k)}$ diverges as an exponentially growing infinite sequence $\zd_2^{(k)}=-\left(G(\vfi^*)\right)^{k-1} \epsilon\cos\varphi^*$. 
The motion never stops and $\zd_2$ cannot be bounded, which is a violation of the FTLS condition. 
\qed

As an example, consider the two-contact configuration given in example 3 in the previous section, whose maps $R$ and $G$ are given in Figure 5(b). The fixed point $\vfi^*=-0.7$ of $R(\vfi)$ satisfies $G(\vfi^*)>1$, hence it is concluded that the equilibrium configuration is unstable, as illustrated in the solution trajectory in Figure 5(e).

An important observation is that Theorem \ref{thm.instability_simple} holds also for non-persistent equilibrium configurations. In this case, contact transitions represented by the dashed edges in the graph of Figure 3 may occur. This implies the possible existence of cyclic paths of contact transitions that do not cross the \Poincare section, making the
map $R(\vfi)$ not well-defined for some values of $\vfi$. Nevertheless, if for some particular value $\vfi^*$, the maps $R$ and $G$ are well-defined and satisfy $\varphi^*=R(\vfi^*)$ and $G(\varphi^*)>1$, then there exists a particular choice of initial condition for which the response grows unbounded, which is sufficient for establishing instability.


The next theorem provides a conservative condition for FTLS.
\begin{thm} \label{thm.stability_simple}
Consider the reduced \Poincare map $R(\vfi)$ and the growth map $G(\vfi)$ associated with the ZOD in the neighborhood of a persistent equilibrium configuration. If $G(\varphi)<1$ for all $\vfi \in I$ where $R(\vfi)$ is defined, then the equilibrium configuration is FTLS.\end{thm}

The core idea of the proof is fairly simple. It is based on the observation that for any initial perturbation lying on the \Poincare section, if the solution path is cyclic and passes through node 5 infinitely many times, then the sequence of collision velocities is bounded by a geometric series as
$|\zd_2^{(k)}| \leq \eta^{k} |\zd_2^{(0)}|$, where $\eta = \max \{G(\vfi): \;\; \vfi \in I\}$. Thus, the solution undergoes a Zeno convergence to a state of two sustained contacts (denoted by DC for double contact) in node 8 or 9 through an infinite sequence of steps that lasts a finite amount of time.
The full proof is rather long and technical since it requires obtaining explicit bounds on solutions that may contain finite or infinite number of mode transitions. Moreover, cases where the initial conditions lie outside the \Poincare section and cases where the angle $\vfi^{(k)}$ approaches the endpoints $\pm \pi/2$ should also be considered. The following lemma contains technical results which are essential for the detailed proof of Theorem \ref{thm.stability_simple}. In particular, it states that the solution reaches a double-contact state (DC) in finite time, and establishes bounds on this time as well as on the divergence of the solution from the original equilibrium at $\vq' \seq 0$:
\begin{lem}
\label{lem:time to Zeno} For a system satisfying the conditions of
Theorem \ref{thm.stability_simple},  
there exist finite positive numbers
$c^{(DC)}$
, $\delta^{(DC)}$
 and $k^{(DC)}$ such that under any given initial state at $t=0$, the system reaches a double-contact state (node 8 or 9 of the transition graph) after a time $t^{(DC)}$ which satisfies the bound
\begin{equation}
t^{(DC)}<c^{(DC)}d(0)\label{eq:lemmastatement for time}.
\end{equation}
Moreover, the pseudometrics $d(t)$ and $D(t)$ along the solution  remain bounded as
\begin{equation}
d(t)<\delta^{(DC)}d(0) \label{eq:lemmastatement for d}
\end{equation}
\begin{equation}
D(t)<k^{(DC)}D(0)\label{eq:lemmastatement for D}
\end{equation}
for all $0\leq t\leq t^{(DC)}$.
\end{lem}
The proof of Lemma \ref{lem:time to Zeno} is given in the Appendix. Using this lemma, the proof of Theorem \ref{thm.stability_simple} can be completed as follows.

\begin{figure*}
\centering{\includegraphics[width=0.7\textwidth]{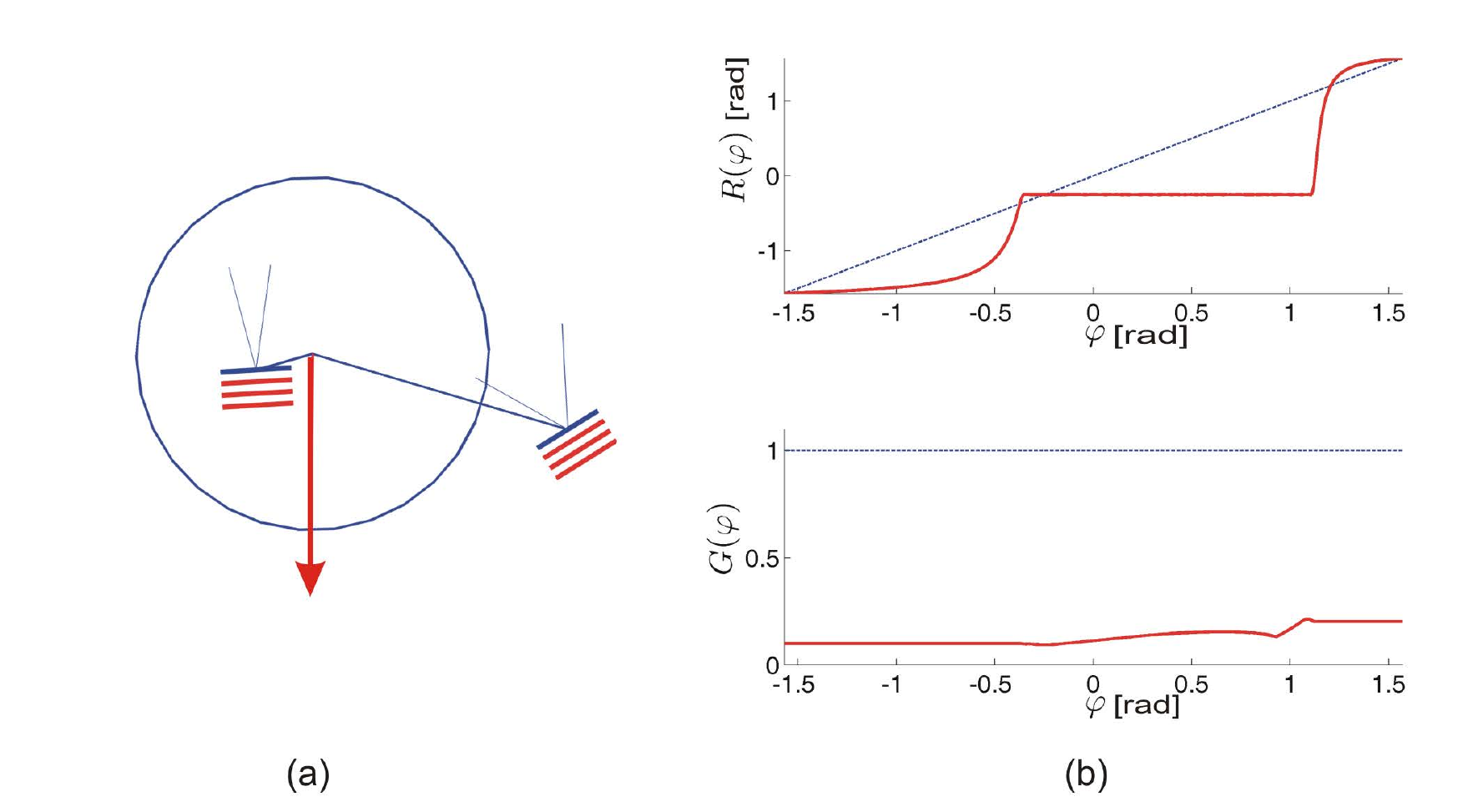}
\caption{Example 4 - (a) Two-contact equilibrium configuration. (b) Plots of the reduced \Poincare map $R(\vfi)$ (top) and  the growth map $G(\vfi)$ (bottom). Theorem \ref{thm.stability_simple} implies FTLS since $G(\vfi)<0.22$ for all $\vfi \in I$.}} \label{fig.example3}
\end{figure*}

\noindent {\bf Proof of Theorem \ref{thm.stability_simple}:}
According to Lemma \ref{lem:time to Zeno}, the solution reaches a two-contact state in a finite bounded time. This can be the immobile SS mode of node 9, or modes of two-contact slippage in node 8. In the latter case, Lemma \ref{lem:finite-impacts} implies that the slippage motion decelerates and stops within an additional time of
$c_{1}\cdot D^{(DC)}$. Hence, we conclude that the system always
stops within a total time of
\begin{equation}
\begin{array}{lll}
t^{(stop)}&<&c^{(DC)}d(0)+c_{1}k^{(DC)}D(0) \\
&\leq&\left(c^{(DC)}+c_{1}\cdot k^{(DC)}\right)\Delta(0)
\end{array}
\label{eq:FTLS-02}
\end{equation}
Meanwhile, $D(t)$ remains bounded by
\begin{equation}
D(t)\leq k^{(DC)}D(0)\leq k^{(DC)}\Delta(0)\label{eq:FTLS-01}
\end{equation}
during the entire motion, including two-contact slippage. The last remaining task is to establish an appropriate upper bound of $|x_{2}(t)|$:
\begin{eqnarray}
|x_{2}(t)| & \leq & |x_{2}(0)|+\intop_{0}^{t}|\dot{x}_{2}(\theta)|d\theta\nonumber \\
 & \leq & \Delta(0)^{2}+\intop_{0}^{t}D(\theta)d\theta\nonumber \\
 & \leq & \Delta(0)^{2}+\underset{t}{\max}D(t)\cdot t^{(stop)}\nonumber \\
 & \leq & \left(1+k^{(DC)}\left(c^{(DC)}+c_{1}\cdot k^{(DC)}\right)\right)\Delta(0)^{2}\label{eq:FTLS-03}
\end{eqnarray}
According to \eqref{eq:FTLS-02}, \eqref{eq:FTLS-01} and \eqref{eq:FTLS-03}, the solution satisfies the FTLS conditions with any arbitrarily small $\epsilon>0$ for any initial condition satisfying $\Delta(0)<\delta$, where
\beq{bounds_FTLS}
\delta = \epsilon \cdot \min \left\{
\begin{array}{c}
\left(c^{(DC)}+c_{1}\cdot k^{(DC)}\right)^{-1}\\
\left(k^{(DC)}\right)^{-1}\\
\left(1+k^{(DC)}\left(c^{(DC)}+c_{1}\cdot k^{(DC)}\right)\right)^{-1/2}
\end{array}
\right \}.
\eeq
Thus, the equilibrium configuration possesses FTLS. 
\qed

\noindent {\bf Example 4 - conservative stability conditions:}  In example 4, the contact configuration is shown in Figure 6(a), and can be verified as a persistent equilibrium. The data of the contacts and external force are given by $\Ag \seq 10^\circ$, $l_1 \seq 1.5$, $l_2 \seq -0.3$, $h \seq 0.15$, $\phi_1 \seq 42.5^\circ$, $\phi_2 \seq 14.2^\circ$, $\mu_1 \seq 0.55$ and $\mu_2 \seq 0.2$. 
According to \eq{bound.painleve}, the conditions for avoiding \Painleve paradox are $\mu_1 < 1.77$ and $\mu_2<47.01$, which are both satisfied.
Plots of the two maps $R(\vfi)$ (top) and $G(\vfi)$ (bottom) are shown in Figure 6(b). It can be verified that $G(\vfi) \leq 0.22$ for all $\vfi \in I$. Therefore, Theorem \ref{thm.stability_simple} implies that this equilibrium configuration is finite-time Lyapunov stable.

\subsection{Generalized stability conditions using the interval graph of $R$}
We now present the most general result of this paper: a stability criterion for almost any two-contact configuration of persistent equilibrium. This condition can be used for analyzing stability of general cases which do not satisfy the conservative theorems of stability or instability, i.e. where the growth map $G(\vfi)$ is not everywhere less than 1 as in Example 4 in Figure 6, and there is no fixed point $\vfi^*=R(\vfi^*)$ with $G(\vfi^*)>1$ as in Example 3 in Figure 5. As a preparatory step, we introduce the notion of {\em interval graph} of the reduced \Poincare map $R(\vfi)$, which is explained as follows. Consider a partition of the interval $I=(-\tfrac{\pi}{2},\tfrac{\pi}{2})$ into $n$ consecutive sub-intervals $I_1 \ldots I_r$ by choosing a series of values $ -\tfrac{\pi}{2} < \vfi_1 < \vfi_2 \ldots \vfi_{r-1} < \tfrac{\pi}{2}$. Each sub-interval is then defined as the closed segment $I_i=[\vfi_{i-1},\vfi_{i}]$ for $i\seq 2 \ldots r-1$, while the first and last sub-intervals, called {\em extremal intervals} are open-ended: $I_1 = (-\tfrac{\pi}{2},\vfi_1]$ and $I_r=[\vfi_{r-1},\tfrac{\pi}{2})$. The interval graph of the reduced \Poincare map $R$ is a directed graph whose vertices are the sub-intervals $I_1 \ldots I_r$. A directed edge $I_j \to I_k$ exists in the graph if there exist $\vfi' \in I_j$ and $\vfi'' \in I_k$ such that $R(\vfi') = \vfi''$. Note that if $R(\vfi)$ is undefined on an entire sub-interval $I_j$ then the corresponding vertex may be a sink of the interval graph. By definition, the interval graph does not account for self-edges $I_j \to I_j$ even in cases where the image of $I_j$ under the map $R(\vfi)$ intersects with $I_j$. A (simple) directed cycle in the interval graph is a path $\{I_{i_1},I_{i_2}, \ldots I_{i_m}\}$ such that all edges $I_{i_j} \to I_{i_{j+1}}$ exist in the interval graph for $j \seq 1 \ldots m-1$ and $i_1 \seq i_m$, while all other pairs $j \neq k$ satisfy $i_j \neq i_k$. The sub-intervals $I_1 \ldots I_r$ are further classified into two categories: {\em safe and unsafe intervals},
such that a sub-interval $I_j$ is safe if for all $\vfi \in I_j$, we either have $G(\vfi)< 1$ or $R(\vfi)$ is undefined. Intervals, which are not safe are classified as unsafe. While the choice of the partition $I_1 \ldots I_r$ and its associated interval graph is arbitrary, FTLS stability is proven here only for the case where there exists a partition satisfying particular properties, which are defined as follows:
\begin{defn} \label{def:safepartition}
For a given two-contact configuration of persistent equilibrium and its associated reduced \Poincare map $R$ and growth map $G$, a partition $I_1 \ldots I_r$ is called a {\em \bf stable partition} if it satisfies the following conditions:
\begin{enumerate}

\item  If the growth map $G$ is defined at an endpoint $\vfi \to \pm \tfrac{\pi}{2}$, then $G(\vfi)$ attains a constant value of $G^-$ or $G^+$ along the corresponding extremal interval, $I_1$ or $I_r$.

\item If the growth map $G(\vfi)$ is defined at an endpoint $\vfi \to \pm \tfrac{\pi}{2}$, then its corresponding value satisfies  $G^{\pm}\neq 1$.


\item {$R(\varphi)-\varphi$ has the same sign (either strictly positive or strictly negative) within each individual unsafe interval.}

\item Directed cycles in the interval graph induced by the map $R(\vfi)$ do not contain unsafe intervals.
\end{enumerate}
\end{defn}
Note that condition 1 is achievable according to the properties of $G$ proven in Lemma \ref{lem:properties_RG}. Additionally, condition 2 is almost always satisfied, except for non-generic contact geometries. The following theorem states that existence of a stable partition implies Lyapunov stability.

\begin{thm} \label{thm.stability_general}
Consider a two-contact persistent equilibrium configuration under the ZOD. If there exists a stable partition of $I$, then the equilibrium configuration is FTLS.\end{thm}
The relations between Theorem \ref{thm.stability_general} and the two previous theorems \ref{thm.instability_simple} and \ref{thm.stability_simple} are as follows. First, if there exists a fixed point $\vfi^*=R(\vfi^*)$ with $G(\vfi^*)>1$ as described in Theorem \ref{thm.instability_simple}, it is obviously impossible construct a stable partition due to violation of condition 3. Second, if the stability conditions of Theorem \ref{thm.stability_simple} are satisfied ($G(\vfi)<1$ for all $\vfi$ where $R(\vfi)$ is defined) then it is trivial to construct a stable partition of $I$ which may contain up to three intervals, two extremal ones and one regular which are all safe, without any unsafe intervals. Thus, Theorem \ref{thm.stability_general} is a direct generalization of Theorem \ref{thm.stability_simple}, and is consistent with Theorem \ref{thm.instability_simple}.

The idea of the proof of  Theorem \ref{thm.stability_general} is that any cyclic path in the transition graph contains only safe intervals and hence it satisfies $G<1$ at each step, which implies Zeno convergence of $\zd_2^{(k)}$ to zero, and thus reaching a state of two sustained contacts within a finite bounded time. A self edge through an unsafe interval $I_u \to I_u$ is not counted as a cyclic path, and it may be repeated only a finite bounded number of times. Therefore, the series $\vfi^{(k)}$ is repelled from unsafe regions where $G>1$ and attracted to cycles with $G<1$. The full rigorous proof is rather technical, and utilizes the following two lemmas, whose proofs appear in the appendix. The first lemma considers the case of unsafe extremal intervals and provides bounds on divergence of solutions that start near the endpoints $\vfi \to \pm \tfrac{\pi}{2}$.

\begin{figure*}
\centering{\includegraphics[width=0.7\textwidth]{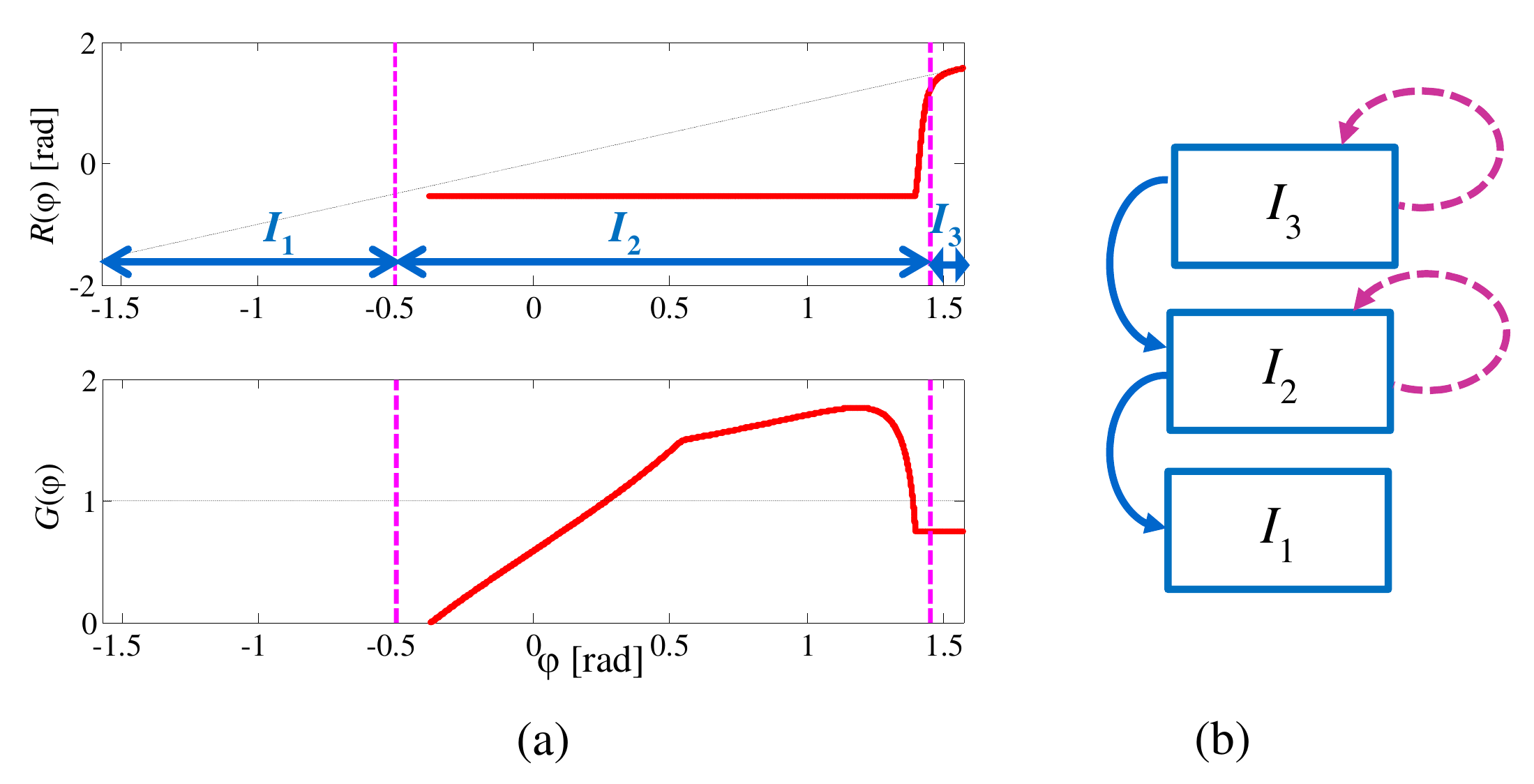}
\caption{Example 5, based on the contact geometry of example 2: (a) Plots of $R(\vfi)$ and $G(\vfi)$ with partition into sub-intervals $I_1$, $I_2$ and $I_3$. (b) The interval graph induced by $R(\vfi)$. Self-edges are denoted by dashed arrows.}} \label{fig.example_interval}
\end{figure*}

\begin{lem}
\label{lem:special intervals} For a two-contact persistent equilibrium configuration, if a stable partition exists and
$G^{+}>1$, ($G^{-}>1$),  then the extremal interval(s) in the partition can be chosen so that
there exist constants $c_{ex}$ and $k_{ex}$ such that any solution that 
    satisfies 
$\varphi^{(m)},\varphi^{(m+1)},...,\varphi^{(m+K)}\in I_{r}$
($I_{1}$),  is bounded by 
\begin{equation}
D^{(i)}\leq k_{ex}\cdot D^{(m)}\label{eq:Dlimit-extremal}
\end{equation}
for all $i=m,m+1,\ldots,m+K+1$, and 
\begin{equation} t^{(m+K+1)}-t^{(m)} 
\leq c_{ex}\cdot D^{(m)}\label{eq:time limit-extremal}
\end{equation}

\end{lem}

The next lemma establishes bounds on solutions under any initial condition, given the existence of a stable partition.

\begin{lem}
\label{lem:time to Zeno2} Consider a two-contact persistent equilibrium configuration under the ZOD. If there exists a stable partition of $I$, then there exist finite positive numbers
$c^{(DC)}$, $\delta^{(DC)}$ and $k^{(DC)}$ such that under any given initial state at $t=0$,  
the system reaches a double-contact state (node 8 or 9 of the transition graph) after a time $t^{(DC)}$ which satisfies the bound
\begin{equation}
t^{(DC)}<c^{(DC)}D(0)\label{eq:lemmastatement for time2}.
\end{equation}
Moreover, the pseudometric 
$D(t)$  remains bounded as
\begin{equation}
 D(t)<k^{(DC)}D(0)\label{eq:lemmastatement for D2}
\end{equation}
for all $0\leq t\leq t^{(DC)}$.
\end{lem}
Note that this lemma is a slightly weaker version of Lemma \ref{lem:time to Zeno}, where \eqref{eq:lemmastatement for D}  remains unchanged; $d(0)$ is replaced with $D(0)$ in the bound \eqref{eq:lemmastatement for time}; and a weaker version of \eqref{eq:lemmastatement for d}:
\begin{equation}
d(t)<\delta^{(DC)}D(0) \label{eq:lemmastatement for d2}
\end{equation}
follows trivially from \eqref{eq:lemmastatement for D2}.


Using this Lemma, the proof of Theorem \ref{thm.stability_general} is almost identical to the proof of Theorem \ref{thm.stability_simple}, where the only change is replacing $d(0)$ with $D(0)$ in \eqref{eq:FTLS-02}, \eqref{eq:FTLS-01} and \eqref{eq:FTLS-03}. Both pseudometrics are anyway bounded by $\Delta(0)$ in these equations, thus the same bounds on $\delta$ in \eq{bounds_FTLS} are obtained in order to establish FTLS.
\qed

\noindent {\bf Example 5 - general stability conditions:} In order to demonstrate the application of Theorem \ref{thm.stability_general} for proving stability, we revisit the contact configuration of example 2 in Figure 4(a) with its associated reduction maps $R(\vfi)$ and $G(\vfi)$ in Figure 4(b). This case is not covered by the simple instability and stability theorems \ref{thm.instability_simple} and \ref{thm.stability_simple}. Using the notion of interval graph, a stable partition into only {\em three} sub-intervals is given by $\vfi_1=-0.5,\;\vfi_2=1.45$, and the sub-intervals are shown on the plots of $R$ and $G$ in Figure 7(a). The intervals $I_1$ and $I_3$ are safe, while only the interval $I_2$ is unsafe. Figure 7(b) shows the interval graph, which contains no cycles (self-edges, appearing in dashed lines, are not counted as cycles). Moreover, the unsafe interval $I_2$ for which $G>1$ maps to the interval $I_1$ for which $R(\vfi)$ is undefined. This corresponds to reaching a two-contact impact followed by finite-time convergence to contact mode SS of static equilibrium. Similar behavior of finite-length paths applies for almost any initial conditions, except for a narrow range where $\vfi^{(1)}>1.53$, for which the series $\vfi^{(k)}$ converges asymptotically to the endpoint $\tfrac{\pi}{2}$. This corresponds to Zeno convergence to PP mode of two-contact slippage, followed by finite-time transition to SS. 

\begin{figure*}
\centering{\includegraphics[width=0.6\textwidth]{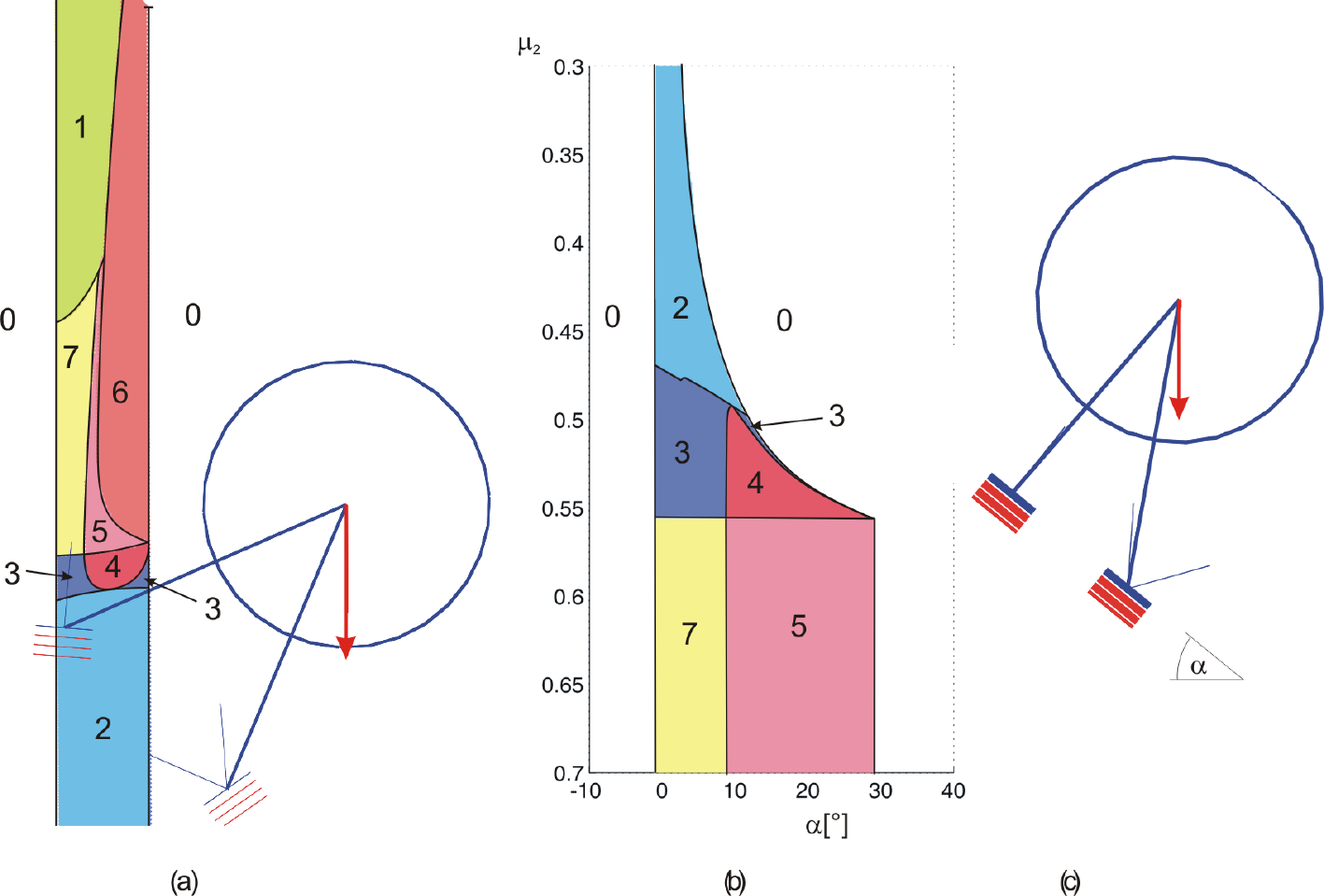} 
\caption{(a) Example 6 -  regions of center-of-mass position with different stability properties plotted over a two-contact configuration. (b) Example 7 - regions with different stability properties in $(\alpha,\mu_2)$ plane. (c) Example 7 - a nominal two-contact configuration. }} \label{fig.stability.regions}
\end{figure*}

\subsection{Stability regions in 2D parameter spaces}
This section is concluded by showing two examples where we plot regions of stability and instability in the plane of two variable parameters that describe the contact configuration.

\noindent {\bf Example 6 - Varying the center of mass position:} we revisit the same reference contact configuration from example 3 shown in Figure 5(a). The center of mass position $\vr_c$ is now varied, while the geometry of the contacts remains unchanged. The external force $\vf_{ex}$ acts at the varying position of $\vr_c$, without external torque. This is very similar to the equilibrium postures analyzed in \cite{or&rimon.icra08b}. Using the stability analysis described above, one can go over a discrete grid of center-of-mass positions and classify the stability properties of the corresponding equilibrium configuration. First, a preliminary check is required for identifying and ruling out cases of infeasible equilibrium or \Painleve paradox. Then, it is checked that the equilibrium configuration is persistent, and the maps $R(\vfi)$ and $G(\vfi)$ are computed. Next, the simple conditions for instability (Theorem \ref{thm.instability_simple}) and conservative stability (Theorem \ref{thm.stability_simple}) are checked. If none of the above conditions are satisfied, the interval $I$ is discredited into 300 sub-intervals, which are classified into safe and unsafe. The associated interval graph is then constructed and all possible direct cycles are identified, in order to verify that the chosen partition is stable. The results are shown in Figure 8(a),
which shows regions of the center of mass with different stability properties plotted in overlay on the contact configuration of the body with an arbitrarily chosen position of the center of mass.
First, note that only a bounded vertical strip of center-of-mass positions corresponds to a configuration where static equilibrium is feasible (see \cite{or&rimon.ijrr2006}), and region 0 represents all infeasible equilibrium configurations. The vertical strip is then divided into regions 1 to 7, according to the following classification. Region 1 denotes configurations where some contact-slippage motions suffer from \Painleves paradox, and thus they are excluded from the analysis. Regions 2 and 3 denotes persistent equilibrium configurations which are stable according to Theorems \ref{thm.stability_simple} and \ref{thm.stability_general}, respectively. Region 4 corresponds to persistent equilibrium configurations which are unstable due to reverse chatter according to Theorem \ref{thm.instability_simple}. Region 5 corresponds to non-persistent equilibrium configurations which are still provably unstable according to Theorem \ref{thm.instability_simple} due to existence of a fixed point $\vfi^* \seq R(\vfi^*)$ such that $G(\vfi^*)>1$. Region 6 corresponds to ambiguous equilibrium configurations (with modes FN and NN) which are unstable according to Theorem \ref{thm.strong}. Finally, region 7 corresponds to non-persistent equilibrium configurations where no unstable fixed point of $R$ exists, hence stability in these regions is not decidable according to our present analysis.

\noindent {\bf Example 7 - Varying the slope angle and the friction coefficient:} we consider a rigid body supported by an inclined plane with slope angle $\alpha$ under gravity force acting at the center of mass, as shown in Figure 8(c). This is very similar to the equilibrium configurations considered in \cite{Varkonyi.icra2012}. The data of the contact geometry and external force are given by $l_1 \seq 1$, $l_2 \seq 0.1$, $h \seq 0.4$, $\phi_1 \seq \phi_2 \seq0$ and $\mu_1 \seq 0.45$. The slope angle $\alpha$ and the friction coefficient $\mu_2$ are varying. Figure 8(b) shows regions in the plane of the parameters $\mu_2$ and $\alpha$, enumerated as region 0 to region 7 according to their stability characterization as described above. \Painleve paradox for this example occurs only for $\mu_2>1.1$, and thus region 1 lies outside the range of the plot's axes. Note that in theory, combining Theorems 1-4  do not give exact conditions for determining stability or instability. That is, sharp decision of stability or instability may not be possible for any persistent equilibrium configuration. Nevertheless, in the practical computation of stability characterization in both examples 6 and 7, we could not find any persistent equilibrium configuration whose stability was ``undecidable''. As for non-persistent equilibrium with undecided stability that appear in region 7, numerical simulations indicated that these regions in both examples are practically stable. Nevertheless, a rigorous theoretical analysis of Lyapunov stability for these cases is beyond the scope of this work, as discussed below in the concluding section.

\section{Conclusion}
In this work we have analyzed Lyapunov stability of a planar rigid body with two frictional contacts under ineslastic impacts and a constant external load. Two mechanisms of instability have been analyzed - ambiguous equilibrium and reverse chatter. On the other hand, convergence to static equilibrium can be achieved via a finite path of mode transitions involving a two-contact collision, or via a decaying infinite Zeno sequence of impacts. Focusing on the subclass of persistent equilibrium configurations, we have studied the zero-order approximation of the dynamics and introduced a two-contact \Poincare map which has then been reduced into two semi-analytic scalar functions -- the reduced \Poincare map $R$ and the growth map $G$, that together encode the system's response to small state perturbations in all directions. Then we have presented conservative theorems for stability and instability, as well as a more general stability analysis based on the interval graph structure of the reduced \Poincare map. The results were demonstrated by showing regions of stability and instability in two-dimensional planes of parameters that describe equilibrium configurations.

We now briefly discuss limitations of our work and sketch possible directions for its future extension. First, the main limitation of our work is imposed by the choice to focus on persistent equilibria. The key difficulty in proving Lyapunov stability for non-persistent equilibrium configurations stems from the fact that any decaying Zeno sequence that converges to a state with two sustained contacts can then be followed by contact separation and transition back to nodes 5 or 6 via the dashed edges in the graph of Figure 3, rather than unique continuation to node 8 followed by stopping at a nearby equilibrium in finite time. Thus, extension of the analysis beyond persistent equilibrium configurations must involve convergence conditions handling ``Zeno sequences of Zeno sequences''. This poses some challenging issues that are currently under investigation.

The second limitation is our assumption of completely inelastic impacts. Frictional impact laws with partial elasticity as manifested in various definitions of restitution coefficients can be found in the literature (cf. \cite{chatterjee&ruina_1998,stronge_book,mason&wang_impact}), though their extension to two or more contacts is much more complicated \cite{brogliato,glocker1995multiple,ivanov1997impact}. In any case, such impacts will substantially  complicate the transition graph in Figure 3, since almost any collision will result in a transition back to node 1 of two free contacts, and the chosen \Poincare section $\Ss$ will probably become useless. Another possible extension of the work is to consider also cases where \Painleve paradox occurs. While cases of inconsistency are typically resolved through well-defined ``tangential impacts'' \cite{champneys&varkonyi_survey2016}, the cases of solution indeterminacy will necessitate incorporating the notion of {\em multi-valued solutions} into the analysis, making it even more challenging.

A fundamentally different yet important direction for extension of the results is validation of the stability analysis by conducting simple experiments of perturbing a rigid body from variable two-contact configurations. A preliminary demonstration of the reverse chatter instability via a diverging sequence of impacts has been shown in an unpublished experiment of a rigid ``biped'' on a slope in a follow-up work of \cite{Varkonyi.icra2012}. Nevertheless, a more systematic setup of experiments has not yet been conducted, and remains as an important future challenge. Finally, the work could also be extended towards stability investigation of a robotic multibody system with multiple degrees of freedom and several contacts. This extension should probably be along the lines of the computational method presented in \cite{posa.stability.tac2016} and oriented towards control, yet it is hoped that it can exploit some intuition and guidelines from the insights gained in our simple low-dimensional work.




\section*{Appendix - Proofs of Technical Details}
\noindent {\bf Proof of Theorem \ref{thm.strong}: } 

Assume that the non-static contact mode XY is consistent at the
equilibrium state $\vq \seq \vqd \seq 0$. Let $\As \subset \real^{6}$
be the set of all states $(\vq,\vqd)$ under which the contact mode XY is kinematically admissible (column 3 of Table 1). For a given metric $\Delta$, define a closed ball of radius $\epsilon$ as:
\[B(\epsilon)= \{(\vq,\vqd) \in \real^6: \;\; \Delta(\vq,\vqd) \leq \epsilon \}. \]
If the contact mode XY is not FF, then it involves contact types S,P or N. The contact mode is consistent at $(\vq,\vqd)=(0,0)$ where  inequality constraints on contact forces for contacts with S,P or N are satisfied. Due to continuity of the dynamic solution in \eq{dyn} and \eq{kin_a} with respect to $\vq$ and $\vqd$, extreme value theorem (EVT) implies the existence of $\epsilon_f>0$ such that the contact forces are consistent under the dynamics of XY for all $(\vq,\vqd) \in B(\epsilon_f)$. In case where the non-static contact mode XY involves separation of the $i^{th}$ contact (F), the normal acceleration $\zdd_i$ evaluated at $(\vq,\vqd)=(0,0)$ under the dynamics of the mode XY must be positive. Using the same continuity and EVT arguments, there also exist $\epsilon_z>0$ and $a_z>0$ such that $\zdd(\vq,\vqd) \geq a_z$ under the dynamics of XY for all $(\vq,\vqd) \in B(\epsilon_z)$. In case where the contact mode XY involves slippage of the $i^{th}$ contact (F or N), the tangential acceleration satisfies $\pm \xdd_i>0$ when evaluated at $(\vq,\vqd)=(0,0)$ under the dynamics of the mode XY, with the $\pm$ sign consistent with F or N. Using the same continuity and EVT arguments, there also exist $\epsilon_x>0$ and $a_x>0$ such that $\pm \xdd(\vq,\vqd)>0 $ and $|\xdd(\vq,\vqd)|\geq a_x$ under the dynamics of mode XY for all $(\vq,\vqd) \in B(\epsilon_x)$. Next, we choose $\epsilon_m=\min\{\epsilon_f,\epsilon_z,\epsilon_x\}$ among the values which are relevant to the mode XY, and then define $\Omega=B(\epsilon_m)$. By construction, the intersection $\Omega \cap \As$ is nonempty and contains states where the contact mode XY is consistent, including the equilibrium state $(\vq,\vqd)=(0,0)$.

Consider now a solution $\vq(t)$ under any initial condition within  $\Omega \cap \As$. Then there exists a finite time $t_f>0$ for which the solution satisfies $(\vq(t),\vqd(t))\in \Omega$ for all $t \in [0,t_f]$. In case where the mode XY involves separation (F) at the $i^{th}$ contact, the solution must satisfy $\zd_i(t) \geq \zd_i(0)+a_z t$ for all $t \in [0,t_f]$. In case where the mode XY involves slippage (P or N) at the $i^{th}$ contact, the solution must satisfy $|\xd_i(t)| \geq |\xd_i(0)|+a_x t$ for all $t \in [0,t_f]$.
In both cases, the solution cannot be bounded within any arbitrarily small neighborhood of $(\vq,\vqd)=(0,0)$ and convergence back to static equilibrium where $z_i=\zd_i=0$ cannot be attained at any  arbitrarily small finite time by setting the initial conditions sufficiently small, which is a violation of the FTLS condition.
\qed

\noindent {\bf Proof of Lemma \ref{lem:finite-impacts}: }

Impacts are considered first, and the existence of $k_1$ satisfying (\ref{eq:impact-Ddlimit}) is proved. After that, values of $c_1$ and $k_1$ satisfying (\ref{eq:lemma-Ddlimit}), \eqref{eq:lemma-tlimit for D}, \eqref{eq:lemma-tlimit for d} are found for each contact mode. The largest one of the candidate values of $c_1$, $k_1$ over all possible contact modes satisfies all conditions of the lemma.

\begin{enumerate}
\item \textbf{Impacts (IF, FI, II):}
the position of the body and
thus $x_i$ and $z_i$ do not vary during an impact. The variations
of the contact velocities during the impact at time $t_{1}$ are
determined by a piecewise linear impact map
\beq{impactlin}
\dot\vq'(t_1^+)=\mathbf{A}\dot \vq'(t_1^-)
\eeq{}

with different impact matrices $\mathbf{A}$ for each impact mode (one or two contacts; slipping or sticking).
%
%
Let  $A$ be the largest absolute value among all elements of the matrices $\mathbf{A}$ over all possible impact modes. Then
\beq{velolin}
\begin{array}{l}
|\dot{x}_2(t_1^+)|,|\dot{z}_1(t_1^+)|,|\dot{z}_2(t_1^+)|
...\\
\;\;\;\;\;\leq
A|\dot{x}_2(t_1^-)|+A|\dot{z}_1(t_1^-)|+A|\dot{z}_2(t_1^-)|)\\
\;\;\;\;\;\leq 3AD(t_1^-)
\end{array}
\eeq
implying the existence of $k_1$ satisfying (\ref{eq:impact-Ddlimit}) for the pseudometric $D$.


Proving the bound (\ref{eq:impact-Ddlimit}) for the pseudometric $d$ goes as follows.
For II impacts, $d(t_1^+)=0$, hence (\ref{eq:impact-Ddlimit}) is satisfied.
Consider now a slipping impact at the single contact $\vr_j$ (FI or IF). The impact matrices of slipping impacts take the form
$$
\mathbf{A}=
\left[
\begin{array}{ccc}
*&*&0\\
*&*&0\\
*&*&1
\end{array}
\right]
$$
because the contact impulse of a slipping impact does not depend on the tangential velocity of any of the two potential contact points. Consequently,
\beq{linveloslip}
\dot{z}_1(t_1^+),\dot{z}_2(t_1^+)
\leq A|\dot{z}_1(t_1^-)|+A|\dot{z}_2(t_1^-)|)\leq 2Ad(t_1^-)
\eeq
implying   the existence of $k_1$ satisfying (\ref{eq:impact-Ddlimit}) for the pseudometric $d$.
To deal with a sticking impact at contact 2,  we exploit the fact that if a planar body undergoes an impact in a state
 not suffering from Painlev\'e's paradox and the impact angle is sufficiently shallow (i.e. $\vfi$ is close to $\pm\pi/2$) then a slipping impact occurs (see also the proof of equation \eq{R.pi2} in Lemma  \ref{lem:properties_RG} for a more detailed discussion of shallow impacts). Specifically, we assume that a sticking impact occurs only if $|\varphi|<\varphi^*$ for some $\varphi^*<\pi/2$.
Thus, we can write similarly to \eq{linveloslip} for $j=1$ and $2$:
\begin{equation}
\begin{array}{rl}
\dot{z}_j(t_1^+)
\leq & A|\dot{x}_2(t_1^-)|+A|\dot{z}_1(t_1^-)|+A|\dot{z}_2(t_1^-)|)
\\
\leq&
A|\dot{z}_1(t_1^-)|+A|\dot{z}_2(t_1^-)|
+A\tan\varphi^*|\dot{z}_2(t_1^-)|\\
\leq& (2+\tan\varphi_0)Ad(t_1^-)
\end{array}
\end{equation}
implying (\ref{eq:impact-Ddlimit}) for the pseudometric $d$.
In the case of a sticking impact at contact 1, we define the impact angle  $\zeta$ as $\zeta=\arctan(\dot x_1(t_1^-)/|\dot z_1(t_1^-)|)$. As in the previous case, a sticking impact implies $|\zeta|<\zeta^*$ for some $\zeta^*<\pi/2$. We also
redefine the impact matrix in a slightly different way:
\beq{impactlinstick}
\dot\vq'(t_1^+)=\vqd'(t_1^-)+\mathbf{A}\left[\dot z_1(t_1^-) \; \dot x_1(t_1^-) \right]^T
\eeq{}
These new definitions allow us to obtain the upper bounds
\begin{eqnarray*}
\dot{z}_1(t_1^+),\dot{z}_2(t_1^+)
&\leq &
d(t_1^-)+A|\dot z_1(t_1^-)|+A|\dot x_1(t_1^-|)\\
&\leq &
d(t_1^-)+A(1+\tan\zeta^*)|\dot z_1(t_1^-)|\\
&\leq &
(1+A(1+\tan\zeta^*))d(t_1^-)
\end{eqnarray*}
where $A$ is again the absolute value of the largest element of $\mathbf{A}$. The last bound implies the existence of $k_1$ satisfying (\ref{eq:impact-Ddlimit}) for the pseudometric $d$
\item \textbf{FF mode:} if the body is in FF mode at $t=t_{1}$, contact 1 accelerates
towards the contact surface by (\ref{eq.ineq_z1dd}). As we show,
a collision occurs at contact 1 after a bounded time, unless
a collision at contact 2 occurs even earlier. Let $t_{1a}=t_{1}$ if $\dot{z}_{1}(t_{1})\leq0$
and $t_{1a}=t_{1}+2\dot{z}_{1}(t_{1})/|\ddot{z}{}_{1}^{(FF)}|$ otherwise.
This choice of $t_{1a}$ means that $z_{1}(t_{1a})=z_{1}(t_{1})$
and $\dot{z}_{1}(t_{1a})=-|\dot{z}_{1}(t_{1})|\leq0$. Because of the
negative sign of $\dot{z}_{1}(t_{1a})$, contact 1 hits the surface  at
\begin{equation}
\begin{array}{lll}
t_{2}&\leq& t_{1a}+\sqrt{\frac{2z_{1}(t_{1a})}{|\ddot{z}{}_{1}^{(FF)}|}}\\
&\leq & t_{1}+\frac{2|\dot{z}_{1}(t_{1})|}{|\ddot{z}{}_{1}^{(FF)}|}+\sqrt{\frac{2z_{1}(t_{1})}{|\ddot{z}_{1}^{(FF)}|}}\\
&\leq& t_{1}+\left(\frac{2}{|\ddot{z}{}_{1}^{(FF)}|}+\sqrt{\frac{2}{|\ddot{z}_{1}^{(FF)}|}}\right)d(t_{1}^-)\\
&\overset{\triangle}{=}&c_{1}^{(FF)}d(t_{1}^-)
\end{array}
\label{eq:t2}
\end{equation}
which gives a valid candidate for $c_1$ in \eqref{eq:lemma-tlimit for d} and also in
(\ref{eq:lemma-tlimit for D}). In the time interval $(t_{1},t_{2})$
, $\sqrt{z_i(t)}$ are bounded by
\begin{equation}
\begin{array}{l}
\sqrt{z_{i}(t)}  \leq...\\
 \;\;\leq  \sqrt{z_i(t_{1})+|\dot{z}_{i}(t_{1})|(t_{2}-t_{1})+\frac{1}{2}|\dot{z}_{i}^{(FF)}|(t_{2}-t_{1})^{2}} \\
  \;\;\leq  \sqrt{d(t_{1})^{2}+d(t_{1})\cdot c_{1}^{(FF)}d(t_{1})+\frac{1}{2}|\dot{z}_{i}^{(FF)}|(c_{1}^{(FF)}d(t_{1}))^{2}}\\
  \;\; =  \sqrt{1+c_{1}^{(FF)}+\frac{1}{2}|\dot{z}_{i}^{(FF)}|c_{1}^{FF\,2}}\, d(t_{1}^.)\\
   \;\; \overset{\triangle}{=}  k_{1a}^{(FF)}d(t_{1}^-).
 \end{array}\label{eq:sqrtzi}
\end{equation}
and the contact velocities $|\dot{x}_{i}|$ and $|\dot{z}_{i}|$ are
bounded by
\begin{equation}
\begin{array}{lll}
|\dot{z}_{i}(t)| & \leq & |\dot{z}_{i}(t_{1})|+|\dot{z}_{i}^{(FF)}|(t_{2}-t_{1})\\
&\leq&\left(1+|\dot{z}_{i}^{(FF)}|c_{1}^{(FF)}\right)d(t_{1}^-)\\
&\overset{\triangle}{=}&k_{1b}^{(FF)}d(t_{1}^-)
\end{array}\label{eq:absdotzi}\\
\end{equation}
\begin{equation}
\begin{array}{lll}
|\dot{x}_{i}(t)| & \leq & |\dot{x}_{i}(t_{1})|+|\dot{x}_{i}^{(FF)}|(t_{2}-t_{1})\\
&\leq&\left(1+|\dot{x}_{i}^{(FF)}|c_{1}^{(FF)}\right)D(t_{1})\\&\overset{\triangle}{=}&k_{1c}^{(FF)}D(t_{1}^-)\end{array}\label{eq:absdotxi}
\end{equation}
Then, $\max \{k_{1a}^{(FF)},k_{1b}^{(FF)},k_{1c}^{(FF)}\}$ is a candidate of $k_1$ in the bounds on both $D$ and $d$ in (\ref{eq:lemma-Ddlimit}).
\item \textbf{FS, SF modes:} if the system is in contact mode XY at $t=t_{1}$
with contact $j$ in F mode and the other contact $i$ in S
mode then the unambigiouity of the examined equilibrium implies $\ddot{z}_{j}^{(XY)}<0$,
i.e. contact point $j$ accelerates toward the contact surface. This inequality can be
used in the same way as \eqref{eq.ineq_z1dd} was used in the case
of the FF mode, to obtain candidate values of $c_1$ and $k_1$.
\item \textbf{PF, NF, FP, FN modes: let $i$} denote the slipping and $j$
the free contact. According to the definition of persistent equilibria
(item 2 in Definition \ref{def:persistence}),
{} $\ddot{z}_{j}^{(XY)}<0$, where XY is the name of the mode in question,
and we can obtain candidate values of $c_1$ and $k_1$ in a way analogous to the previous case of FF mode.
\item \textbf{PP, NN, PN, NP modes: }The unambiguity of the equilibrium
implies that two-contact slipping is decelerating under the ZOD,
{} and slipping stops at time $t=t_{1}+{\dot x_{i}}(t_{1})/\ddot{x}{}_{i}^{(XY)}|$
 $\leq t_{1}+|\ddot{x}{}_{i}^{(XY)}|^{-1}D(t_{1})$,
yielding the candidate value $c_1^{(XY)}=|\ddot{x}{}_{i}^{(XY)}|^{-1}$ in (\ref{eq:lemma-tlimit for D}). At the same time, $d(t)=0$
is constant and $D(t)=|\dot{x}_2(t)|$ is a decreasing function
of time, hence $k_1=1$ is a valid candidate for eq. (\ref{eq:lemma-Ddlimit}
\item \textbf{SS mode:} the task becomes trivial as $D(t)=d(t)=0$ so that $c_1$,$k_1$ can be chosen as 1.
\end{enumerate}
\qed

\noindent {\bf Proof of Lemma} \ref{lem:continuity_RG}:

Recall that there are two possible scenarios for discontinuity of the maps $R$ and/or $G$:
\begin{enumerate}
\item discontinuity of \eq{varphi} when $\zd_2^{(k+1)}=0$
\item discontinuity of the full \Poincare map $\vP$
\end{enumerate}
Scenario 1 occurs at a transition point from single impact to double impact (II), therefore it is a point where $R$ and $G$ become undefined rather than a point of discontinuity. 

In order to investigate scenario 2, we will decompose $\vP$ into four maps. There are two possible paths in the transition graph from the \Poincare section to itself: $4\rightarrow 3\rightarrow 2\rightarrow 5$ or $4\rightarrow 6\rightarrow 2\rightarrow 5$. Accordingly, we consider $\vP=\vP_5 \circ \vP_2 \circ \vP_{3} \circ \vP_4$ or $\vP=\vP_5 \circ \vP_2 \circ \vP_{6} \circ \vP_4$, each corresponding to a mapping of state variables from reaching a given node of the transition graph until leaving it. The choice between the two paths depends on parameteres of the system, but it is not state-dependent (i.e. switching between the two paths as the initial conditions are varied is not possible). It is easy to show that the impact maps $\vP_2$ and $\vP_4$ are continuous and piecewise-smooth. Nonsmoothness is due to the possibility of switching between sticking and slipping impacts, nevertheless the contact impulses change continuously at the transition from stick to slip. The remaining maps $\vP_3$, $\vP_5$ and $\vP_6$ are induced by a piecewise smooth continuous-time dynamics (non-smoothness is again caused by the possibility of slip-stick and slip reversal) crossing a contact surface in state space ($z_1=0$ or $z_2=0$) and thus they may become discontinuous due to a grazing bifurcation (i.e. when a small change in initial conditions causes a nearly tangential crossing of $z_i=0$ that did not exist before). Nevertheless, it can be shown that grazing bifurcations are impossible in our system under the ZOD. The proof of this statement is presented for $\vP_3$. This map is defined by the relation
$$
(x_2^{**},\xd_2^{**},\zd_1^{**},z_2^{**},\zd_2^*)=\vP_3(x_2^*,\zd_2^*,\xd_2^*)
$$
where $^{*}$ denotes values of state variables immediately after an FI impact (where the remaining state variables are 0: $z_1^* = \zd_1^* = z_2^*=0$) and $^{**}$ denotes values of state variables immediately before the subsequent IF impact (where we always have $z_1^{**}=0$). A grazing bifurcation would correspond to $\zd_1^{**}=0,\ddot{z}_1^{**} > 0$ nevertheless this is impossible due to $\eq{ineq_z1dd}$. Similar reasoning (skipped for brevity) reveals that  $\vP_5$ and $\vP_6$ are continuous for any non-ambiguous equilibrium under the ZOD.

\noindent {\bf Proof of equation \eq{R.pi2} in Lemma  \ref{lem:properties_RG}: }

The case of  $\vfi$ being close to $+\pi/2$ is considered in detail. The proof for $\vfi\rightarrow-\pi/2$ is completely analogous. Due to the scaling invariance (\ref{eq.inv_scale}), we may assume pre-impact velocity $\dot{x}_{2}^{(k)}=+1$
at the time of crossing the Poincar\'{e} section. Then, by definition
of $\vfi$, one obtains $\underset{\vfi\rightarrow\pi/2}{\lim}\dot{z}_{2}^{(k)}=0$.
A sufficiently shallow impact ($\vfi$ close to $\pi/2$)  implies a slipping impact with vanishing contact impulse: $\underset{\vfi\rightarrow\pi/2}{\lim}\hat{\mathbf{f}}_{1},\hat{\mathbf{f}}_{2}=0$
leaving the contact velocities unchanged by the impact event: $\underset{\vfi\rightarrow\pi/2}{\lim}\dot{x}_{2}^{+}=1$
and $\underset{\vfi\rightarrow\pi/2}{\lim}\dot{z}_{1}^{+},\dot{z}_{2}^{+}=0$
where $^{+}$ means post-collision velocities immediately after
crossing the Poincar\'{e} section.
(The implication of a slipping impact is not valid for systems subject to the \Painleve paradox, which may undergo sticking 'tangential impacts' \cite{champneys&varkonyi_survey2016}. Nevertheless, such systems are not examined here).

After the slipping impact, the motion continues with
PF or FF mode (nodes 3,6 of the transition graph), which ends when
contact 1 hits the contact surface. Due to the vanishing normal velocity $\dot{z}_{1}^{+}$
and to (\ref{eq.ineq_z1dd}), the duration of this phase of the motion vanishes in
the limit: $\underset{\vfi\rightarrow\pi/2}{\lim}t_{node\,3,6}=0$,
and thus the velocities again remain unchanged. It can be shown similarly that another slipping impact with vanishing contact impulse and another phase of slipping motion occurs before the system returns to the Poincar\'{e} section, hence
\begin{equation}
\underset{\vfi\rightarrow\pi/2}{\lim}\dot{z}_{2}^{(k+1)}=0\label{eq:lemma1-01}
\end{equation}
and
\begin{equation}
\underset{\vfi\rightarrow\pi/2}{\lim}\dot{x}_{2}^{(k+1)}=1\label{eq:lemma1-02}
\end{equation}
This implies that $\lim_{\vfi\to\pi/2} R(\vfi) =\pi/2$, which proves \eq{R.pi2}.
\qed

\noindent {\bf Proof of equation \eq{G.pi2} in Lemma  \ref{lem:properties_RG}: }

We have proved in the previous point (presenting the proof of equation \eq{R.pi2}) that for $\vfi$ sufficiently close to $\pi/2$ (or $-\pi/2$), the Poincar\'{e}-cycle consists of slipping impacts and slipping motion in the positive (or negative) direction. In this situation, the invariance relation \eq{G-endpoints-sliding}  implies a constant value of G, proving \eq{G.pi2}.
\qed

\noindent {\bf Proof of equation \eq{dR.pi2} in Lemma  \ref{lem:properties_RG}: }
Again, we investigate the case $\vfi\rightarrow\pi/2$ in detail. Similarly to the previous point, we assume
$\dot{x}_{2}^{(k)}=+1$ to be a constant,
 and $\dot{z}_{2}^{(k)}$ will be used as independent variable instead of $\varphi$.  The definitions of $R$ and $\varphi$ are used to express $R'$ as
\begin{equation}
\begin{array}{lll}
\frac{dR(\vfi)}{d\vfi}&=&
\frac{d\left[\arccot(\dot{z}_{2}^{(k+1)}/\dot{x}_{2}^{(k+1)})\right]}
{d\left[\arccot(\dot{z}_{2}^{(k)}/\dot{x}_{2}^{(k)})\right]}\\
&=&
\frac{d\left[\arccot(\dot{z}_{2}^{(k+1)}/\dot{x}_{2}^{(k+1)})\right]/{d\dot{z}_{2}^{(k)}}}{d\left[\arccot(\dot{z}_{2}^{(k)}/\dot{x}_{2}^{(k)})\right]/{d\dot{z}_{2}^{(k)}}}
\end{array}\label{eq:R-derivative-01}
\end{equation}
Since $G=G^+$ in
the range of our interest according to  \eq{G.pi2}, we have $\dot{z}_{2}^{(k+1)}=G^{+}\dot{z}_{2}^{(k)}$, yielding
\begin{equation}
\begin{array}{lll}
\frac{dR(\vfi)}{d\vfi}&=&
\frac{d\left[\arccot(G^{+}\dot{z}_{2}^{(k)}/\dot{x}_{2}^{(k+1)}\right]/{d\dot{z}_{2}^{(k)}}}
{d\left[\arccot(\dot{z}_{2}^{(k)})\right]/{d\dot{z}_{2}^{(k)}}}\\
&=&
\frac{
\left(1+\left( G^{+}\dot{z}_{2}^{(k)}/\dot{x}_{2}^{(k+1)}\right)^{2}\right)^{-1}
}
{\left(1+\left( \dot{z}_{2}^{(k)}\right)^2\right)^{-1}}
\cdot G^{+}\cdot\\
&&...\left(\dot{x}_{2}^{(k+1)}-\dot{z}_{2}^{(k)}\frac{d\dot{x}_{2}^{(k+1)}}{d\dot{z}_{2}^{(k)}}\right)\left(\dot{x}_{2}^{(k+1)}\right)^{-2}
\end{array}\label{eq:R-derivative-02}
\end{equation}
Eq. (\ref{eq:lemma1-01}) and (\ref{eq:lemma1-02}) yield
\begin{equation}
\underset{\vfi\rightarrow\pi/2}{\lim}\frac{dR(\vfi)}{d\vfi}=\frac{\left(1+0\right)^{-1}}{\left(1+0\right)^{-1}}G^{+}\left(1-0\right)1^{-2}=G^{+}\label{eq:R-derivative-03}
\end{equation}
\qed



\noindent {\bf Proof of equations (\ref{eq:lemmastatement for time}) and (\ref{eq:lemmastatement for d}) in Lemma \ref{lem:time to Zeno}: }
First, we prove that $G(\vfi)$ has a well-defined maximum.  $G$ is a continuous function, which is defined over a collection of open subintervals of the $(-\pi/2,\pi/2)$ interval. Its domain is bounded by the points $\varphi=\pm\pi/2$ and by points where $G$ becomes undefined due to the occurrance of a two-contact impact. $G$ has well-defined limit values at both types of limit points, namely $G^\pm$ at $\pm\pi/2$ and 0 at points of transition to a two-contact impact. Thus, the domain of $G$ can be extended to obtain a continuous function over a compact, closed set. The extreme value theorem ensures that $G$ has a well-defind maximum value $\eta$. Furthermore $\eta<1$ by the conditions of Theorem \ref{thm.stability_simple}. The rest of the proof is divided into two cases and makes extensive use of this property of $G$.

\textbf{Case 1 - crossing the Poincar\'{e} section infinitely many times:}
if the trajectory of the system crosses the Poincar\'{e} section infinitely
many times, the proof goes as follows: inspection of the transition
graph reveals that along any path terminating at the Poincar\'{e} section,
but not crossing it earlier, the number of impacts plus the number
of pieces of trajectory without contact mode transitions is not more
than $7$. According to bounds (\ref{eq:impact-Ddlimit}) and (\ref{eq:lemma-Ddlimit})
of Lemma \ref{lem:finite-impacts}, the pseudometric $d$ remains bounded
until reaching the Poincar\'{e} section for the first time at $t=t^{(1)}$ by
\begin{equation}
d(t)\leq k_{1}^{7}d(0)\,\, for\,\, t<t^{(1)}\label{eq:dlimit 0--t1}
\end{equation}
 and the time of reaching the Poincar\'{e} section is bounded by
\[
t^{(1)}\leq c_{1}\left(1+k_{1}+...+k_{1}^{6}\right)d(0)
\]
Analogously, between the first and second events of crossing the Poincar\'{e}
section, the number of impacts plus the number of pieces of trajectory
without contact mode transitions is not more than 6, implying
\begin{equation}
d(t)\leq k_{1}^{6}d^{(1)}\leq k_{1}^{13}d(0)\,\, for\,\, t^{(1)}<t<t^{(2)}\label{eq:dlimit t1---t2}
\end{equation}
\begin{equation}
\begin{array}{lll}
t^{(2)}-t^{(1)}&\leq& c_{1}\left(1+k_{1}+...+k_{1}^{5}\right)d^{(1)}\\
&\leq& c_{1}\left(1+k_{1}+...+k_{1}^{5}\right)k_{1}^{7}d(0)
\end{array}\label{eq:t(2) limit}
\end{equation}

The $G$ map measures the growth of $d$ during a \linebreak Poincar\'e cycle
 according to \eq{dD} and \eq{G}. The bound $\eta$ of $G$ yields
\begin{equation}
d{}^{(2)}\leq \eta d{}^{(1)}\leq \eta k_{1}^{7}d(0)\label{eq:d(2) limit}
\end{equation}
and by induction,
\begin{equation}
d{}^{(n)}\leq \eta^{n-1}k_{1}^{7}d(0)\label{eq:d(q) limit}
\end{equation}
for any integer $n$. The last equation implies analogues of (\ref{eq:dlimit t1---t2}), (\ref{eq:t(2) limit}):

\begin{equation}
d(t)\leq k_{1}^{6}d^{(n)}\leq \eta^{n-1}k_{1}^{13}d(0)\,\, for\,\, t^{(n)}<t<t^{(n+1)}\label{eq:dlimit tq-1   tq}
\end{equation}
\[
t^{(n+1)}-t^{(n)}\leq c_{1}\left(1+k_{1}+...+k_{1}^{5}\right)\eta^{n-1}k_{1}^{7}d(0)
\]
for any integer $n$. Hence,$\underset{n\rightarrow\infty}{\lim}d^{(n)}=0$
and infinitely many impacts occur in finite time (Zeno point), where
the body eventually transitions into double-contact state. Since $\eta<1$, the total
time required to reach DC  is bounded by
\begin{equation}
\begin{array}{l}
t^{(DC)}=\underset{n\rightarrow\infty}{\lim}t^{(n)} \\
\;\;\; \leq  \left[\left(1+k_{1}+...+k_{1}^{5}\right)k_{1}^{7}\left(1+\eta+\eta^{2}+...\right)+
\right.\\\left.
\;\;\;\;\;\;\;\;\;\;\;\;...\left(1+k_{1}...k_{1}^{6}\right)\right]c_{1}d(0)\\
\;\;\; =  \left[\left(1+k_{1}+...+k_{1}^{5}\right)k_{1}^{7}\left(1-\eta\right)^{-1}+
\right.\\ \left.
\;\;\;\;\;\;\;\;\;\;\;\;
...\left(1+k_{1}...k_{1}^{6}\right)\right]c_{1}d(0)\\
 \;\;\; \overset{\triangle}{=}  c^{(DC)}d(0)
 \end{array}\label{eq:tZeno}
\end{equation}
which proves (\ref{eq:lemmastatement for time}), whereas (\ref{eq:dlimit 0--t1})
and (\ref{eq:dlimit tq-1   tq}) prove (\ref{eq:lemmastatement for d}).

\textbf{Case 2 - crossing the Poincar\'{e} section a finite number of
times: }the Poincar\'{e} section crossed a finite number of times,
if the system undergoes a two-contact impact during the $\chi^{th}$ cycle,
where $\chi$ is an arbitrary integer. $\chi=0$ means that the Poincar\'{e}
section is not crossed at all. After the two-contact impact, the system
arrives to a double contact state immediately. The upper bound for
$d(t)$ found in case 1 remains valid. The bound (\ref{eq:tZeno}) of $t^{(DC)}$ obtained as the sum of an infinite geometric series yields a conservative bound in case 2, because $t^{(DC)}$ is also bounded by a finite part of this series.


\noindent {\bf Proof of equation (\ref{eq:lemmastatement for D}) in Lemma \ref{lem:time to Zeno}: }
Even though $R$ is undefined at $\pm\pi/2$, it has well-defined derivatives at these points. First we define a bound of its second derivatives at the endpoints:
\beq{rho}
\rho=
1+\max\left\lbrace
0,\lim_{\vfi\to\pi/2}d^2R/d\vfi^2,-\lim_{\vfi\to-\pi/2}d^2R/d\vfi^2
\right\rbrace
\eeq
Next, we choose a scalar $\varphi_{cr}$ which is sufficiently close to $\pi/2$ to satisfy
\begin{equation}
\pi/2-\vfi_{cr}<\min\left\{\pi/4,\frac{G^\pm}{\rho}\right\}\label{eq:phicr-01}
\end{equation}
furthermore, the following relations are also true for all $|\varphi|>\varphi_{cr}$:
\begin{eqnarray}
G(\varphi)=G^{\pm}\label{eq:phicr-02}\\
\pi/2\mp R(\varphi)>
G^{\pm}(\pi/2\mp\varphi)-\frac{1}{2}\rho(\pi/2\mp\varphi)^2\label{eq:phicr-03}\\
\begin{cases}
\pi/2\mp R(\varphi)<
\frac{1+G^{\pm}}{2}
\left(\pi/2\mp\varphi\right)\; if \; G^{\pm}<1 \\
\pi/2\mp R(\varphi)>
\frac{1+G^{\pm}}{2}
\left(\pi/2\mp\varphi\right)\; if \; G^{\pm}>1
\end{cases}
\label{eq:phicr-04}
\end{eqnarray}
In all of these relations, the upper member of the pairs $\pm$, $\mp$ should be considered if $\vfi>0$ and the lower one in the opposite case. The polynomial bounds (\ref{eq:phicr-03}), (\ref{eq:phicr-04}) are  satisfied if $\vfi$ is sufficiently close to $\pm\pi/2$ according to \eq{R.pi2}, \eq{dR.pi2} and \eq{rho}.
The rest of the  proof is divided into three cases.

\textbf{Case 1: the Poincar\'{e} section is not crossed during the motion} inspection of the transition graph reveals that along
any path reaching double-contact state via node 7, but not crossing
the Poincar\'{e} section, the number of impacts plus the number of pieces
of trajectory without contact mode transitions is not more than $5$.
Hence, statements (\ref{eq:impact-Ddlimit}) and (\ref{eq:lemma-Ddlimit}) of
Lemma \ref{lem:finite-impacts} imply $D(t)\leq k_{1}^{5}\cdot D(0)$.

\textbf{Case 2: the Poincar\'{e} section is crossed finitely or infinitely many times, and $|\varphi^{(n)}|<\varphi_{cr}$ for
all $n$.} By using the upper bounds of $d(t)$ given by (\ref{eq:lemmastatement for d}) and by noting that the definition of the pseudometrics $D$, $d$ implies
\beq{D/d ratio1}
D^{(n)}/d^{(n)} = 1 \;if\;|\vfi^{(n)}|<\pi/4
\eeq
\beq{D/d ratio2}
D^{(n)}/d^{(n)} = |\tan(\vfi^{(n)})| \;if\;|\vfi^{(n)}|\geq \pi/4
\eeq
 whenever the system crosses the Poincar\'{e} section, we obtain
\begin{equation}
D^{(n)}\leq\tan\varphi_{cr}\cdot\delta^{(DC)}d(0)\label{eq:D^(q) for phi<phi_cr}
\end{equation}
whereas Lemma \ref{lem:finite-impacts}
yields the bound
\begin{equation}
D(t)\leq k_{1}^{5}D^{(n)}\leq\tan\varphi_{cr}k_{1}^{5}\delta^{(DC)}d(0)\label{eq:Dlimit between crossings}
\end{equation}
 for all $t^{(n+1)}>t>t^{(n)}$ and for all $n$, completing the proof.
If the Poincar\'{e} section is crossed finitely many times due to a two-contact impact, (\ref{eq:Dlimit between crossings}) is also valid for the
time interval starting at the last crossing of the Poincar\'{e} section,
and ending at $t^{(DC)}$, which completes the proof. Notice that
$d(t^{(DC)})=0$ and $D(t^{(DC)})\leq\tan\varphi_{cr}\cdot d(t^{(DC)})=0$,
i.e. the system stops immediately after the two-contact impact.

\textbf{Case 3: there exists $n$ for which $|\varphi^{(n)}|\geq\varphi_{cr}$.} We discuss the case of crossing the Poincar\'{e} section with $+\varphi^{(n)}>\varphi_{cr}$. The case of $-\varphi^{(n)}>\varphi_{cr}$ is analogous. Now, since $\vfi^{(\nu+1)}=R(\vfi^{(\nu)})$, (\ref{eq:phicr-04}) implies that ...
$\varphi^{(n)}$,$\varphi^{(n+1)}$,$\varphi^{(n+2)}$,...
is an increasing sequence inside the $(\varphi_{cr},\pi/2)$ interval. Let $K$ denote the index of the first crossing
with angle $\varphi>\varphi_{cr}$ .

If $K=1$, then the system arrives to the Poincar\'{e} section for the
first time, along a path where the total number of impacts and epsiodes
of motion free of contact mode transitions is not more than 7. Hence,
\begin{equation}
D^{(K)}\leq k_{1}^{7}D(0)\,\label{eq:D(k) bound for K=00003D1}
\end{equation}

In the case of $K>1$, the bound (\ref{eq:D^(q) for phi<phi_cr})
found in Case 2 is valid for all $n<K$. During one cycle of crossing
the Poincar\'{e} section, the number of impacts and the number of pieces
of trajectory without impacts and contact mode transitions is at most
6. Thus,
\begin{equation}
D^{(K)}\leq k_{1}^{6}D^{(K-1)}\leq k_{1}^{6}\tan\varphi_{cr}\delta^{(DC)}d(0)\label{eq:D(k) bound}
\end{equation}
 by Lemma \ref{lem:finite-impacts}. Bounds on the sequence $D^{(n)}$ for $n>K$
are obtained by a recursive formula as follows. Eq.(\ref{eq:phicr-01}), (\ref{eq:phicr-02}), \eq{D/d ratio2} and the following bounds of the cotangent function
\beq{}
\pi/2-x<\cot(x)<(\pi/2-x)+(\pi/2-x)^2\; if\; \pi/4<x<\pi/2
\eeq
together imply for all $1\leq n\in\mathbb{N}$,
\begin{equation}
\begin{array}{lll}
D^{(K+n)} & = & D^{(K+n-1)\cdot}\frac{d^{(K+n)}}{d^{(K+n-1)}}\frac{\cot\varphi^{(K+n-1)}}{\cot\varphi^{(K+n)}}\\
 & = & D^{(K+n-1)}G(\varphi^{(K+n-1)})\frac{\cot\varphi^{(K+n-1)}}{\cot R(\varphi^{(K+n-1)})}\\
 & \leq & D^{(K+n-1)}G^{+}\cdot\frac{\pi/2-\varphi^{(K+n-1)}+\left(\pi/2-\varphi^{(K+n-1)}\right)^{2}}{\pi/2-R(\varphi^{(K+n-1)})}
 \end{array}
\end{equation}
Next we use (\ref{eq:phicr-03}) to obtain
\begin{equation}
\begin{array}{l}
D^{(K+n)} < \\ \;\; D^{(K+n-1)}G^{+}\frac{\pi/2-\varphi^{(K+n-1)}+\left(\pi/2-\varphi^{(K+n-1)}\right)^{2}}
{G^{\pm}(\pm\pi/2-\varphi^{(K+n-1)})-\frac{1}{2}\rho(\pm\pi/2-\varphi^{(K+n-1)})^2}\\
 \;\; = D^{(K+n-1)}\frac{1+\pi/2-\varphi^{(K+n-1)}}
 {1-\frac{\rho}{2G^+}\left( \pi/2-\varphi^{(K+n-1)} \right)}
\end{array}
\end{equation}
By exploiting (\ref{eq:phicr-01}) together with the relation
\beq{}
\frac{1+c_1}{1-c_2} < 1+2 c_1+2 c_2 \; if \; 0<c_1 \;\; 0<c_2<1/2
\eeq
we obtain
\begin{equation}
 D^{(K+n)} <  D^{(K+n-1)}
 \left[
 1\! +\! \left(2\! +\!\frac{\rho}{G^+} \right)\left(\pi/2\! -\!\varphi^{(K+n-1)}\right)\right]
 \label{eq:Drecursive}
\end{equation}
At the same time, (\ref{eq:phicr-04}) implies that $\vfi$ converges to $\pi/2$:
\begin{equation}
\begin{array}{lll}
\pi/2-\varphi^{(K+n-1)}&\leq&\left(\pi/2-\varphi^{(K)}\right)\left(\frac{1+G^{+}}{2}\right)^{n-1}\\
&\leq&\left(\pi/2-\varphi_{cr}\right)\left(\frac{1+G^{+}}{2}\right)^{n-1}
\end{array}\label{eq:varphi convergence}
\end{equation}
Eq. (\ref{eq:Drecursive}) and (\ref{eq:varphi convergence}) yield
\begin{equation}
\begin{array}{l}
D^{(K+n)} \leq ...\\
\;\;\; ...D^{(K+n-1)}\left(1+\left(2+\frac{\rho}{G^+} \right)\left(\pi/2-\varphi_{cr}\right)\left(\frac{1+G^{+}}{2}\right)^{n-1}\right)\label{eq:Drecursive-2}
\end{array}
\end{equation}
The recursive formula (\ref{eq:Drecursive-2}) allows one to obtain
a global upper bound of $D^{(K+n)}$ for all $n$:
\begin{equation}
\begin{array}{l}
D^{(K+n)} \leq... \\
\;\leq D^{(K)}\prod_{\nu=1}^{n}\left(1 \! + \! \left(\pi/2 \! - \! \varphi_{cr}\right)\left(2 \! + \! \frac{\rho}{G^+} \right)\left(\frac{1 \! + \! G^{+}}{2}\right)^{\nu-1}\right)\\
 \;\leq  D^{(K)}\prod_{\nu=1}^{\infty}\left(1 \! + \! \left(\pi/2 \! - \! \varphi_{cr}\right)\left(2 \! + \! \frac{\rho}{G^+} \right)\left(\frac{1 \! + \! G^{+}}{2}\right)^{\nu-1}\right)
 \label{eq:D(K) as a limit}\\
 \; \overset{\triangle}{=}  D^{(K)}l(\varphi_{cr},G^{+}).
\end{array}
\end{equation}
Notice that the infinite product represented by $l(\varphi_{cr},G^{+})$ is of the
type $\prod_{\nu=1}^{\infty}\left(1+\gamma_1{\gamma_2}^{\nu}\right)$ with $\gamma_2<1$,
and it has a well-defined finite, positive limit value. Then, we use (\ref{eq:D^(q) for phi<phi_cr})
(valid for $n<K$), as well as (\ref{eq:D(k) bound for K=00003D1}) and (\ref{eq:D(k) bound})
to conclude that
\begin{eqnarray}
D^{(n)} & \leq & \max\left\{\begin{array}{c}
k_{1}^{6}\tan\varphi_{cr}\delta^{(DC)}\\
k_{1}^{7}
\end{array}\right\}l(\varphi_{cr},G^{+})\cdot d(0)\label{eq:D(K+n)}
\end{eqnarray}
Finally, the bounded number of contact mode transitions and impacts
during a cycle of crossing the Poincar\'{e} section implies
\begin{equation}
\begin{array}{lll}
D(t) & \leq&\underset{n}{k_{1}^{6}\max}D^{(n)}\\
 & \leq & k_{1}^{6}\max\left\{\begin{array}{c}
k_{1}^{6}\tan\varphi_{cr}\delta^{(DC)}\\
k_{1}^{7}
\end{array}\right\}l(\varphi_{cr},G^{+})\cdot d(0) \\
&\overset{\triangle}{=}& k^{(DC)}d(0)\label{eq:Dlimit-final}
\end{array}
\end{equation}
for all $t\leq t^{(DC)}$, which proves (\ref{eq:lemmastatement for D}).

If the Poincar\'{e} section is crossed a finite number of times, then
the infinite product in (\ref{eq:D(K) as a limit}) can be replaced
by a finite product, which yields a sharper bound. Nevertheless (\ref{eq:Dlimit-final}) remains valid.

\noindent {\bf Proof of 
Lemma \ref{lem:special intervals}}

Previously, we have investigated  the convergence of the sequence $\varphi^{(n)}$ to $\pi/2$ in the case of $G^+<1$, as part of the proof of equation (\ref{eq:lemmastatement for D}) in Lemma \ref{lem:time to Zeno}.  Here, we investigate the divergence of $\vfi^{(n)}$ from $\pi/2$ inside the extremal interval $I_r$ in the case of $G^+>1$ in a similar fashion.

We have defined a scalar $\vfi_{cr}$, which satisfies the bounds (\ref{eq:phicr-01})-(\ref{eq:phicr-04}).
If $\vfi_{cr}$ is contained in $I_r$, then we divide $I_r$ at $\vfi_{cr}$ into two sub-intervals: a regular unsafe interval $I_{r*}$ and an extremal unsafe interval $I_{r}$. It can be demonstrated easily that none of the conditions of a stable partition is violated by this modification. Now, $\vfi>\vfi_{cr}$ in $I_r$ and thus the exponential divergence of $\vfi^{(n)}$ from $\pi/2$ is dictated by the relation (\ref{eq:phicr-04}), from which one can deduce a bound analogous to (\ref{eq:varphi convergence}) for all $0\leq n \leq K$:
\begin{equation}
\begin{array}{lll}
\pi/2-\varphi^{(m+K-n)}&\leq&\left(\pi/2-\varphi^{(m+K)}\right)\left(\frac{1+G^{+}}{2}\right)^{-n}\\
&\leq&\left(\pi/2-\varphi_{cr}\right)\left(\frac{1+G^{+}}{2}\right)^{-n}
\end{array}\label{eq:varphi divergence}
\end{equation}
One can also derive an analogue of the bound (\ref{eq:Drecursive})  in the present situation for all $0\leq n \leq K$:
\begin{equation}
\begin{array}{l}
D^{(m+K-n+1)}  \leq...\\
\;\;\;...  D^{(m+K-n)}\left(1+
\left(\pi/2-\varphi_{cr}\right)
\left(2+\frac{\rho}{G^+} \right)\left(\frac{1+G^{+}}{2}\right)^{-n)}\right)
\end{array}\label{eq:Drecursive-2a}
\end{equation}
The recursive formula (\ref{eq:Drecursive-2a}) allows one to obtain
a global upper bound of $D^{(m+K-n+1)}$ for all $0\leq n \leq K$:
\begin{equation}
\begin{array}{l}
D^{(m+K-n+1)}   \leq ...\\
\;\;\leq D^{(m)}\prod_{\nu=-n+1}^{K-1}\left(1+\left(\pi/2-\varphi_{cr}\right)
\left(2+\frac{\rho}{G^+} \right)
\left(\frac{2}{1+G^{+}}\right)^{\nu}\right)\\
\;\;\leq
D^{(m)}\prod_{\nu=1}^{\infty}\left(1+\left(\pi/2-\varphi_{cr}\right)
\left(2+\frac{\rho}{G^+} \right)
\left(\frac{2}{1+G^{+}}\right)^{\nu}\right)\\
\;\; \overset{\triangle}{=}  D^{(m)}l(\varphi_{cr},G^{+}).
 \end{array}\label{eq:D(K) as a limita}
\end{equation}
The infinite product $l(\varphi_{cr},G^{+})$ is of the
type
$$\prod_{n=1}^{\infty}\left(1+\gamma_1\cdot \gamma_2^{n}\right)$$ with $\gamma_2<1$,
and it has a well-defined finite, positive limit value, which  can be chosen as the bounding constant $k_{ex}$ in (\ref{eq:Dlimit-extremal}).

The constant bound (\ref{eq:D(K) as a limita}) of $D$ and eq. (\ref{eq:varphi divergence}) together ensure that the finite sequence $d^{(m+K)}$, $d^{(m+K-1)}$,...,$d^{(m)}$ is bounded from above by a geometric series with exponent $<1$. It can be proved then by using Lemma \ref{lem:finite-impacts} that $t^{(m+K+1)}-t^{(m+K)}$, $t^{(m+K)}-t^{(m+K-1)}$,...,$t^{(m+1)}-t^{(m)}$ are also bounded from above by a finite geometric series in which the sum of the terms remains bounded. This gives a bound $c_{ex}$ for proving (\ref{eq:time limit-extremal}). The rest of the proof is very similar to the proof of (\ref{eq:lemmastatement for time}) in Lemma \ref{lem:time to Zeno}, and thus the details are not repeated here. The lower endpoint $\vfi \to -\pi/2$ is treated analogously.

\noindent {\bf Proof of Lemma \ref{lem:time to Zeno2}}

First, we prove (\ref{eq:lemmastatement for d2}) via combining the proof of (\ref{eq:lemmastatement for d}) in Lemma \ref{lem:time to Zeno} with Lemma \ref{lem:special intervals}.
According to the definition of safe partitions, if the sequence $\varphi^{(n)}$ $n=1,2,...$ includes elements belonging
to an extremal unsafe interval, then these
elements must form a single finite block of adjacent elements: ($\varphi^{(m)}$,$\varphi^{(m+1)}$,...,$\varphi^{(m+K)}$) where $K$ can be arbitrarily large.
Furthermore, Lemma \ref{lem:special intervals} establishes the bound
\begin{equation}
d^{(m+K+1)}\leq D^{(m+K+1)}\leq k_{extr}D^{(m)}\label{eq:bound01}
\end{equation}
Let $\Upsilon_1,\Upsilon_2,...,\Upsilon_u$ denote the regular unsafe intervals. According to condition 3 of Definition \ref{def:safepartition}, those elements within the sequence $\varphi^{(n)}$,
which belong to  $\Upsilon_{i}$,
form a subsequence composed of adjacent elements, say $\varphi^{(a)}$,$\varphi^{(a+1)}$,$\varphi^{(a+2)}$,..,$\varphi^{(a+b-1)}$.
According to condition 2, this subsequence is monotonic.

It can be shown using the extreme value theorem, that the continuous function $|R(\vfi)-\vfi|$ has a strictly positive lower bound
$\beta$ over regular unsafe intervals. It follows then that the number $b$ of elements in the subsequence
$$\varphi^{(a)},\varphi^{(a+1)},\varphi^{(a+2)},..,\varphi^{(a+b-1)}$$
is bounded by $b\leq L_{i}\beta^{-1}+1$, where $L_{i}$ is the length
of $\Upsilon_{i}$. The total number of elements in the whole sequence
$\varphi^{(n)}$ $n=1,2,...$, belonging to regular, unsafe intervals
is bounded by $K_{ru}\leq\sum_{i=1}^{u}(L_{i}\beta^{-1}+1)$. These intervals partially
cover the $(-\pi/2,\pi/2)$ interval, hence $\sum_{i=1}^{u}L_{i}\leq\pi$,
yielding
\begin{equation}
K_{ru}\leq\pi\beta^{-1}+u\label{eq:bound02}
\end{equation}
For each $n$ such that $\varphi^{(n)}$ is in a regular unsafe interval,
the limited number of impacts and contact mode transitions, and Lemma \ref{lem:finite-impacts} imply
\begin{equation}
d^{(n+1)}\leq k_{1}^{6}d^{(n)}\label{eq:bound03}
\end{equation}
 Finally, for any natural number $Q$,  the number of elements within the sequence $\varphi^{(n)}$
$n=1,2,...,Q$ belonging to safe intervals is bounded from below by
\begin{equation}
K_{s}=Q-K_{ru}-K\geq\max\left\{0,Q-\pi\varepsilon^{-1}-u-K\right\}\label{eq:bound05}
\end{equation}
 where $K$ is the total number of steps in one of the two extremal unsafe interval.
 For each $n$ such that $\varphi^{(n)}$
is in a safe interval, the existence of a maximum value $\eta<1$ of $G$ over all safe intervals (by the extreme value theorem) implies
\begin{equation}
d^{(n+1)}\leq \eta d^{(n)}\label{eq:bound06}
\end{equation}

 \noindent We combine the bounds (\ref{eq:bound01})-(\ref{eq:bound06}) to obtain
\newpage
\begin{equation}
d^{(Q)}\leq\left\{
\begin{array}{l}
k_{extr}D^{(1)}\cdot k_{1}^{6(\pi\varepsilon^{-1}+u)}... \\
\;\;\;\;\;\;\;\;\;\;\;\; \mbox{if}  \,\,\,\, Q\leq\pi\varepsilon^{-1}+u+K \\
k_{extr}D^{(1)}\cdot k_{1}^{6(\pi\varepsilon^{-1}+u)}\cdot \eta^{Q-\pi\varepsilon^{-1}-u-K}... \\
\;\;\;\;\;\;\;\;\;\;\;\;\mbox{if}  \,\,\,\, Q>\pi\varepsilon^{-1}+u+K
\end{array}\right\} \label{eq:dmax}
\end{equation}
whereas $D^{(1)}\leq k_1^7 D(0)$ by Lemma \ref{lem:finite-impacts} .The upper bound found above is an exponentially decreasing function
of $Q$ except for small values of $Q$, where it is constant. This result plays the same role in the present proof as (\ref{eq:d(q) limit}) in the proof of Lemma \ref{lem:time to Zeno}. Hence, the relation
(\ref{eq:lemmastatement for d2}) can be proven analogously to (\ref{eq:lemmastatement for d}) in Lemma \ref{lem:time to Zeno}.

The bound (\ref{eq:lemmastatement for D2}) can be proven as follows: for all $Q$ such that $\vfi^{(Q)}$ is not in an extremal interval, the ratio $D^{(Q)}/d^{(Q)}$ is bounded from above by the relations \eq{D/d ratio1} -\eq{D/d ratio2}, hence (\ref{eq:dmax}) can be used to bound $D^{(Q)}$ from above. For all $Q$ such that $\vfi^{(Q)}$ is in an extremal interval, Lemma \ref{lem:special intervals} provides the necessary bounds on $D^{(Q)}$. These bounds can be combined in a straightforward way to demonstrate (\ref{eq:lemmastatement for D2}). Finally, the time bound (\ref{eq:lemmastatement for time2}) can  be proved by combining (\ref{eq:dmax}) with Lemma \ref{lem:finite-impacts}.


\bibliographystyle{plain}


\end{document}